\documentclass[aip,reprint]{revtex4-1}

\usepackage[utf8]{inputenc}
\usepackage{amsmath,amssymb}
\usepackage{bm}
\usepackage{physics}
\usepackage[version=4]{mhchem}
\usepackage{graphicx}
\usepackage[dvipsnames]{xcolor}
\usepackage{dcolumn}
\usepackage{multirow}
\usepackage{tabularx}
\usepackage{soul}%
\usepackage{xr} % eXternal References
\makeatletter
\newcommand*{\addFileDependency}[1]{%
  \typeout{(#1)}
  \@addtofilelist{#1}
  \IfFileExists{#1}{}{\typeout{No file #1.}}
}
\makeatother
\newcommand*{\myexternaldocument}[1]{%
    \externaldocument{#1}%
    \addFileDependency{#1.tex}%
    \addFileDependency{#1.aux}%
}
\myexternaldocument{waterSI}
\def\AM1W{\mbox{AM1-W}}
\def\PM6fm{\mbox{PM6-fm}}
\def\DFTB2iBi{\mbox{DFTB2-iBi}}
\def\GFNxTB{\mbox{GFN-xTB}}
\def\BLYPD3{\mbox{BLYP-D3}}
\newcommand{\gOO}{\ensuremath{g_{\text{OO}}(r)}}
\newcommand{\gOH}{\ensuremath{g_{\text{OH}}(r)}}
\newcommand{\gHH}{\ensuremath{g_{\text{HH}}(r)}}
\newcommand{\nHB}{\ensuremath{\expval{n_\text{HB}}}}
\newcommand{\Rbeta}{\ensuremath{\qty{R, \beta}}}
\newcommand{\WRbeta}{\ensuremath{W(R, \beta)}}
\newcommand{\fDD}{\ensuremath{f_{\text{DD}}}}
\newcommand{\fSD}{\ensuremath{f_{\text{SD}}}}
\newcommand{\fND}{\ensuremath{f_{\text{ND}}}}
\newcommand{\DHvap}{\ensuremath{\Delta H_\text{vap}}}
\newcommand{\subfig}[3][,]{%
  \setbox1=\hbox{\includegraphics[#1]{#3}}% store subfig in box
  {\centering\usebox1}
  \centerline{\scriptsize #2}
}
\newcommand{\subfigWRBeta}[3][,]{%
  \setbox1=\hbox{\includegraphics[#1]{#3}}% store subfig in box
  {\centering #2 \\[3pt]}
  {\centering\raisebox{6ex}{\rotatebox{90}{\scriptsize$\beta$ (degrees)}}\;\usebox1}
  \centerline{\scriptsize R(\AA{})}
}

\draft % marks overfull lines with a black rule on the right

\begin{document}

\title{%
Benchmarking semi-empirical quantum chemical methods on liquid water
}

\author{Xin Wu}
\author{Hossam Elgabarty}
\affiliation{%
  Dynamics of Condensed Matter and Center for Sustainable Systems Design,
  Chair of Theoretical Chemistry, Paderborn University,
  Warburger Str. 100, D-33098, Paderborn, Germany
}
\author{Vahideh Alizadeh}
\affiliation{%
Mulliken Center for Theoretical Chemistry, Institute for Physical and Theoretical Chemistry, University of Bonn, Beringstr. 4, 53115 Bonn, Germany
}
\affiliation{%
Center for Advanced Systems Understanding, Untermarkt 20, D-02826 G\"orlitz, Germany
}
\affiliation{%
Helmholtz Zentrum Dresden-Rossendorf, Bautzner Landstra{\ss}e 400, D-01328 Dresden, Germany
}
\author{Andrés Henao}
\author{Frederik Zysk}
\affiliation{%
  Dynamics of Condensed Matter and Center for Sustainable Systems Design,
  Chair of Theoretical Chemistry, Paderborn University,
  Warburger Str. 100, D-33098 Paderborn, Germany
}
\author{Christian Plessl}
\affiliation{%
  Department of Computer Science and
  Paderborn Center for Parallel Computing,
  Paderborn University,
  Warburger Str. 100, D-33098 Paderborn, Germany
}
\author{Sebastian Ehlert}
\affiliation{%
  AI for Science, Microsoft Research, Evert van de Beekstraat 354, 1118 CZ Schiphol, The Netherlands 
}
\author{Jürg Hutter}
\affiliation{%
  Physical Chemistry Institute, University of Zurich,
  Winterthurerstrasse 190, CH-8057 Zurich, Switzerland
}
\author{Thomas D. Kühne}
\email{tkuehne@cp2k.org}
\affiliation{%
Center for Advanced Systems Understanding, Untermarkt 20, D-02826 G\"orlitz, Germany
}
\affiliation{%
Helmholtz Zentrum Dresden-Rossendorf, Bautzner Landstra{\ss}e 400, D-01328 Dresden, Germany
}
\affiliation{%
TU Dresden, Institute of Artificial Intelligence, Chair of Computational System Sciences, N\"othnitzer Stra{\ss}e 46 D-01187 Dresden, Germany
}

\date{\today}

\begin{abstract}
  Stimulated by the renewed interest and recent developments in semi-empirical quantum chemical (SQC) methods for noncovalent interactions, we examine the properties of liquid water at ambient conditions by means of molecular dynamics (MD) simulations, both with the conventional NDDO-type (neglect of diatomic differential overlap) methods, \textit{e.g.}\ AM1 and PM6, and with DFTB-type (density-functional tight-binding) methods, \textit{e.g.}\ DFTB2 and \GFNxTB{}. Besides the original parameter sets, some specifically reparametrized SQC methods (denoted as \AM1W{}, \PM6fm{}, and \DFTB2iBi{}) targeting various smaller water systems ranging from molecular clusters to bulk are considered as well. The quality of these different SQC methods for describing liquid water properties at ambient conditions are assessed by comparison to well-established experimental data and also to \BLYPD3{} density functional theory-based \textit{ab initio} MD simulations. Our analyses reveal that static and dynamics properties of bulk water are poorly described by all considered SQC methods with the original parameters, regardless of the underlying theoretical models, with most of the methods suffering from too weak hydrogen bonds and hence predicting a far too fluid water with highly distorted hydrogen bond kinetics. On the other hand, the reparametrized force-matchcd \PM6fm{} method is shown to be able to quantitatively reproduce the static and dynamic features of liquid water, and thus can be used as a computationally efficient alternative to electronic structure-based MD simulations for liquid water that requires extended length and time scales. \DFTB2iBi{} predicts a slightly overstructured water with reduced fluidity, whereas \AM1W{} gives an amorphous ice-like structure for water at ambient conditions.
\end{abstract}


\maketitle

\section{\label{sec:intro}INTRODUCTION}

Since the first molecular dynamics (MD) simulation of liquid water,\cite{Stillinger1971}
it is arguably one of the most widely studied systems due to its anomalous properties and
important role in many chemical and biological processes.\cite{Stillinger1980, eisenberg2005structure, Ball2008}
In order to better understand the many complex behaviors of liquid water,
a full spectrum of theoretical models ranging from
% 1. MM
classical molecular mechanical (MM) force fields\cite{Jorgensen2005, Guillot2002, vega2011simulating} to
% 2. DFT
various density functional theory (DFT),\cite{DFTWater} or
% 3. WFN
even high-level wavefunction-based quantum chemical approaches\cite{MP2Water}
has been developed over the past several decades.

Most of the properties of liquid water at ambient conditions can be accurately predicted
by conventional classical force fields.\cite{TIP3PTIP4P, SPCwater, SPCE, TIP4P2005, spura2015nuclear}
Despite the success of these force fields in reproducing many experiments,
% 1. proton transfer
when chemical reactions are explicitly involved,\cite{Marx2006, markovitch2008special, mundy2009hydroxide, luduena2011mixed, hassanali2013proton, hassanali2014aqueous, andrade2020free, zeng2023mechanistic}
% 2. liquid-vapor interface
different regions of the phase diagram are to be explored,\cite{Mundy2004, goldman2009ab, Kuehne2011JPCL, kessler2015structure, wilhelm2019dynamics, ohto2019accessing, zhang2021phase, palos2024current}
% 3. low-frequency vibrations
or some subtle quantum mechanical effects are mainly concerned,\cite{LowFreq, spura2015fly, rybkin2017nuclear, Elgabarty_TDK_SA2020, andreani2020hydrogen, balos2022time, stolte2024nuclear}
the applicability of these force fields may become quite restricted.
Hence, an explicit quantum mechanical treatment of water is still indispensable.

\textit{Ab initio} molecular dynamics (AIMD),\cite{MarxHutter,Kuehne2014WIRE, hutter2024ab} where 
the forces acting on the nuclei are computed ``on-the-fly'' by accurate electronic structure methods, at least in principle allows to ameliorate 
many of the aforementioned limitations encountered
by classical MD simulations.
Moreover, along with the continuous increase in computing power and
% 1. LS-DFT
the development of so-called linear-scaling DFT methods,\cite{Goedecker1999, richters2014self}
% 2. LS-MD
as well as accelerated AIMD approaches,\cite{CPMD1985, Kuehne2007PRL, niklasson2008extended}
numerous DFT-based simulations of liquid water at various conditions have been reported
in the literature.\cite{ThreeGGA, DFTWaterAssessment, DiffSamp, HybridDF, morrone2008nuclear, Kuehne2009JCTC, distasio2014individual, ojha2024nuclear}
Nevertheless, the computational cost of these AIMD simulations
remains rather high. Thus, they are unsustainable for certain studies that require extensive length and time scales.

Semi-empirical quantum chemical (SQC) methods are a particular kind of
low-cost electronic structure theory.%
\cite{Elstner2011, Thiel2014, bannwarth2021extended}
Like many \textit{ab initio} quantum chemical and DFT approaches,\cite{pople1999nobel, kohn1999nobel} SQC methods 
solve the electronic structure problem in an explicit manner.
However, various approximations, as well as adjustable parameters are introduced
in SQC methods that result in a dramatic increase
in computational efficiency without a significant loss of accuracy,
e.g.\ SQC calculations are observed to be 2--3 orders of magnitude faster than
typical DFT calculations using medium-sized basis sets.
\footnote{To be precise, we consider ``double-$\zeta$ plus polarization'' basis sets
to be medium-size.}
Hence, MD simulations of complex systems demanding long time and length scales
may benefit from SQC methods.

%
% original SQC methods
%
The development of SQC methods has recently come back into the spotlight.
\cite{PM6, PM7, PDDG, NOMNDO, OM1, OM2, OM3, OMx, ODMx,
DFTB2, DFTB3Full, GFN1xTB, GFN2xTB}
The currently most popular SQC methods can be classified into two schemes,
namely NDDO- (neglect of diatomic differential overlap)
and DFTB-type (density-functional tight-binding),
in accordance with the different theoretical formalisms.\cite{MNDO, DFTB1}
% 1. NDDO-type: AM1 and PM6
The former NDDO-type methods, e.g.\ AM1\cite{AM1} and PM6\cite{PM6}, are based on
electronic integral approximations to the underlying Hartree-Fock theory.
% 2. DFTB-type:
By contrast, DFTB-type methods are generally derived from a series expansion of
the DFT energy expression with respect to a reference electron density.
% DFTB2
The rather popular DFTB2 approach, for instance, includes energy contributions up to the second-order term.\cite{DFTB2}
% GFN-xTB
The \GFNxTB{} approach is a new DFTB-type method aiming at yielding good molecular
\underline{G}eometries, vibrational \underline{F}requencies,
and \underline{N}oncovalent interactions with e\underline{x}tensions
within \underline{TB} Hamiltonian and the basis set.\cite{GFN1xTB}
Although it shares some characteristics with DFTB2 and DFTB3,%
\cite{DFTB2, DFTB3Full}
\GFNxTB{} has unique features, %
such as an angular momentum-dependent second-order term to account for charge fluctuations
and the general avoidance of diatomic pairwise parameters.
% GFN2-xTB
Being currently the most sophisticated SQC method in its family, the GFN2-xTB scheme includes anisotropic effects by incorporating multipolar contributions up to
the second-order terms without noticeable increase in computational cost.\cite{GFN2xTB}
Last but not least, both \GFNxTB{} and GFN2-xTB have been parametrized
for all chemical elements in the periodic table up to $Z = 86$.\cite{bannwarth2021extended}

%
% reparametrized SQC methods
%
It has been widely established that many anomalous properties of liquid water
stem from the complex hydrogen bond (H-bond) network formed by individual water molecules
and their neighbors.\cite{Stillinger1980, luzar1996hydrogen, luzar1996effect, li2011quantum, Elgabarty2015NC, richardson2016concerted, clark2019opposing}
Unfortunately, since the early NDDO era, SQC methods have had a poor reputation when it comes to describing
H-bond interactions.\cite{Thiel1988, QC2011}
A vast variety of schemes have been developed over the years to improve the treatment of H-bonding in SQC methods.
\cite{MNDOH, PM3PIF, Truhlar_PMO2, SCPNDDO4water, HBCorr4thGen, DFTB3CPE, DFTB3CPED3}
Another commonly employed strategy is the specific reparametrization with respect to
smaller systems of water.\cite{OMxW, DFTB2iBi, AM1W, DFTB3OBW, PM6fm, DFTB3diagMioiBi}
On the one hand, this approaches can easily be implemented on top of existing SQC methods.
On the other hand, unlike MM-based H-bond corrections,\cite{PM3PIF, PM3MAIS, HBCorr4thGen}
the parameters responsible for the electronic structure can also be refined in this way.
Therefore, it can be expected that specifically reparametrized SQC method able to
improve the accuracy for describing not only water itself 
but also other relevant aqueous systems.
%
% AM1-W, PM6-fm, and DFTB2-iBi
%
Promising reparametrized SQC variants for water of the AM1,\cite{AM1} PM6\cite{PM6} and DFTB2\cite{DFTB2} models are 
\AM1W{},\cite{AM1W} \PM6fm{},\cite{PM6fm} and \DFTB2iBi{}\cite{DFTB2iBi}, respectively. 

There have been several studies focusing on MD simulations of bulk water using either NDDO- or DFTB-type SQC methods.%
\cite{RuizLopez2005, SCPNDDO4water, Thiel2008, OMxW,
DFTB2gammaHB4Water, Voth2010, Voth2013, Voth2014, QC2011, DFTB3OBW}
Nevertheless, it remains essential to conduct a comprehensive benchmark of MD simulations with both NDDO- and DFTB-type SQC methods for
a wide range of static and dynamic properties of liquid water at ambient conditions, using the original, as well as specifically reoptimized parameter sets.
Beside assessing the conceptual strengths and weaknesses of the considered SQC methods, the present systematic study is of great value for further computational simulations of related aqueous systems, such as the liquid/vapor interface, or ``on-water'' catalysis for instance.\cite{karhan2014role, salem2020insight}

\section{\label{sec:mndodftb}SEMI-EMPIRICAL NDDO- AND DFTB-type MODELS}
%%%Bedarf: major revision

Here, we provide a brief side-by-side comparison of the NDDO- and DFTB-type methods to highlight the similarities and differences of these SQC methods (see table \ref{tab:SQC}).
As a starting point, we chose the total energy expressed in an atomic orbital basis given as 
\begin{equation}
    E = \frac{1}{2} \sum_{\mu\nu} (H_{\mu\nu}^\text{core} + F_{\mu\nu}) P_{\mu\nu} + E_{NN},
    \label{total_energy}
\end{equation}
where \(\mu\) and \(\nu\) are the indices of the atomic orbitals, \(H_{\mu\nu}^\text{core}\) is the one-electron Hamiltonian, \(F_{\mu\nu}\) is the Fock matrix, \(P_{\mu\nu}\) is the density matrix and \(E_{NN}\) is the nuclear-nuclear repulsion energy.
The Fock matrix elements are computed as 
\begin{equation}
    F_{\mu\nu} = T_{\mu\nu} + V_{\mu\nu}^\text{Ne} + J_{\mu\nu} + K_{\mu\nu} + V_{\mu\nu}^\text{xc}
\end{equation}
where the kinetic energy \(T_{\mu\nu}\) and the external potential \(V_{\mu\nu}^{Ne}\) are part of \(H_{\mu\nu}^{core}\).
The two-electron interactions split into the Coulomb potential \(J_{\mu\nu}\), non-local exchange potential \(K_{\mu\nu}\) and the semi-local exchange-correlation (XC) potential \(V_{\mu\nu}^{xc}\).
Based on these terms we will classify the potential contributions in the SQC methods.

\begin{table*}[htb]
\caption{Comparison of DFT matrix elements with NDDO-type and DFTB-type methods.}
    \begin{ruledtabular}
    \centering
    \begin{tabularx}{\textwidth}{ccccc}
         &    & DFT  &   NDDO-type  &  DFTB-type  \\[1ex]
         \hline
   \(\begin{aligned}T_{\mu\nu}\end{aligned}\)  & kinetic energy   & 
             \( \begin{aligned}
                 - \frac{1}{2}\langle \psi_\mu \mid \nabla^2  \mid \psi_\nu \rangle 
             \end{aligned}\)  
                        & \(\begin{aligned}U_{\mu\nu} \quad/\quad \beta_\text{AB} S_{\mu\nu}\end{aligned}\)
                        &  \(\begin{aligned}\frac{1}{2}k(\varepsilon^0_\mu + \varepsilon^0_\nu)S_{\mu\nu}\end{aligned}\)  \\[3ex]
  \(\begin{aligned}E_\text{NN}\end{aligned}\)   & N-N repulsion    &
           \(\begin{aligned}
                \frac{1}{2}\sum_{A\neq B} \frac{Z_\text{A} Z_\text{B}}{R_\text{AB}}
           \end{aligned} \) &
            \(\begin{aligned}\frac{1}{2}\sum_{\text{A}\neq \text{B}} Z_\text{A} Z_\text{B} (s^\text{A}s^\text{A}\mid s^\text{B}s^\text{B} ) \end{aligned}\) &
          \(\begin{aligned} E_\text{rep} + \frac{1}{2}\sum_{\mu\nu}\gamma_{\mu\nu}n_{\mu,0}n_{\nu,0} + \frac{1}{3}\sum_\text{A}\Gamma_\text{A}n_\text{A,0}^3\end{aligned}\)    \\[3ex]
  \(\begin{aligned}\hat{V}_\text{Ne} \end{aligned}\)   & N-e attraction   &
           \(\begin{aligned} - \frac{1}{2}\langle \psi_\mu \mid \frac{Z_\text{A}}{|\mathbf R_\text{A} - \mathbf r|}  \mid \psi_\nu \rangle \end{aligned}\) &   
            \(\begin{aligned}\frac{1}{2}\sum_{A} Z_\text{A} (\mu\nu \mid s^\text{A}s^\text{A}) \end{aligned}\) &
             \(\begin{aligned} -\frac{1}{2} S_{\mu\nu}\sum_{\sigma}(\gamma_{\mu\sigma} + \gamma_{\nu\sigma}) n_{\sigma,0} + \frac{1}{2} S_{\mu\nu} (\Gamma_\text{A} n_{\text{A},0}^2 + \Gamma_\text{B} n_{\text{B},0}^2 )  \end{aligned}\)   \\[5ex]
        \(\begin{aligned}J_{\mu\nu} \end{aligned}\)
     & e-e repulsion    
           
        &\(\begin{aligned} \sum_{\lambda \sigma}(\mu\nu \mid \lambda\sigma) P_{\lambda\sigma}  \end{aligned}\) 
        & \(\begin{aligned}  \sum_{\lambda \sigma \in B}(\mu\nu \mid \lambda\sigma) P_{\lambda\sigma} \end{aligned}\)
        & \(\begin{aligned}
            \frac{1}{4}S_{\mu\nu} \sum_{\lambda\sigma} (\gamma_{\mu\sigma} + \gamma_{\mu\lambda} + \gamma_{\nu\lambda} + \gamma_{\nu\sigma} ) S_{\lambda\sigma}P_{\lambda\sigma} \\ - S_{\mu\nu} (\Gamma_A n_{A,0} p_B + \Gamma_B n_{B,0} p_A)
        \end{aligned}
           \)   \\[5ex]
     
        \(\begin{aligned}K_{\mu\nu} \end{aligned}\)
        & Fock exchange    & 
     \(\begin{aligned}\sum_{\lambda \sigma}(\mu\lambda \mid \nu\sigma) P_{\lambda\sigma}  \end{aligned}\) &
     \(\begin{aligned}  \sum_{\substack{\lambda \in A\\\sigma \in B}}(\mu\lambda \mid \nu\sigma) P_{\lambda\sigma}  \end{aligned}\)& 
     --- \\[5ex]
     
     \(\begin{aligned}E_\text{xc} \end{aligned}\)
     &  XC energy       
     & \(\begin{aligned}\int \varepsilon_\text{xc}\left[\rho\right](\mathbf r) \rho(\mathbf r) d\mathbf r\end{aligned}\)    
     &  ---   
     &  \(\begin{aligned} \frac{1}{3}\sum_\text{A} q^3_A\Gamma_A \end{aligned}\)  \\[5ex]
    \end{tabularx}
    \end{ruledtabular}
  %  \caption{Caption}
    \label{tab:SQC}
\end{table*}

The kinetic energy (\(\hat{T}_\text{e}\)) terms mainly drive the covalent bond formation.
In DFT, the integral for the kinetic energy term is solved for all electrons, while only the valance atomic orbitals (\( \psi_\mu \)) are taken into account in SQC methods.
In general, SQC methods neglect most integrals and replace them with parameters to greatly reduce the computational cost.
Thus, by embedding integral contributions into parameters, SQC methods can use simplified Hamiltonian and wave functions.
The kinetic energy term in the core Hamiltonian of SQC methods usually is approximated as a nonlinear function of the overlap integral (\(S_{\mu\nu}\)).
In NDDO-type methods the kinetic energy is approximated using a rescaled overlap integral \(\beta_{AB} S_{\mu\nu}\) for the offsite elements, with \(\beta_{AB}\) as a pair parameter for the atoms A and B.
For the onsite elements, a diagonal matrix \(U_{\mu\nu}\) of the orbital energies \(\varepsilon_\mu^0\) of the free atom is usually used.
In DFTB-type methods, the kinetic energy term is approximated using the extended H\"uckel theory (EHT) approach,\cite{ExtendedHueckelTheory} which approximates the Hamiltonian elements as an average of the orbital energies of the free atoms together with a non-linear scaling factor for the overlap integral. For DFTB-type methods, the energy contributions are expressed in density fluctuations \(p_\mu\), computed from the density matrix by Mulliken charge partitioning, and atomic reference densities \(n_{\mu,0}\).
Depending on the granularity of the parametrization, energies are expressed in orbital-resolved charges \(p_\mu\), shell-resolved charges \(p_\ell\), or atom-resolved charges \(p_\text{A}\).

The nuclear-nuclear repulsion energy (\(E_\text{NN}\)) needs to account for the exchange-repulsion of the core densities in SQC methods, which combines the interaction of the nuclei and the core electrons.
Each SQC method applies different approximations to screen the nuclear charges with the core electrons.
NDDO simply models \(E_\text{NN}\) with an electrostatic repulsion between two overlap charge distributions of two s orbitals centered on atom A and B \((s^As^A\mid s^B s^B)\).
However, in the DFTB and GFN-xTB methods, the nuclear-nuclear Coulomb interactions are modeled as the sum of the parametrized repulsion energy \(E_\text{rep}\), the Coulomb interaction of the reference densities (\(\frac{1}{2}\sum_{\mu\nu}\gamma_{\mu\nu}n_{\mu,0}n_{\nu,0}\)) and the onsite 3rd order contribution of the reference densities (\(\frac{1}{3}\sum_\text{A}\Gamma_\text{A}n_\text{A,0}^3\)).
Therein, \(\Gamma\) is defined as Hubbard derivatives determining how chemical hardness changes with the charge density.\cite{DFTB2,DFTB3Prop}
It is worth mentioning that third-order contributions are not included in all DFTB-type methods such as DFTB2, in which only interactions up to second-order are considered.

The equivalent of the nuclear-electron attraction (\(V^\text{Ne}_{\mu\nu}\)) in the NDDO methods is modeled with two-center two-electron integrals, involving an overlap charge distribution of an s orbital centered on each atom.
In the DFTB and GFN-xTB methods, the nuclear-electron attraction is described by the Coulomb potential of the reference density.
It acts via a Coulomb kernel \(\gamma\) on the charge fluctuations expressed as Mulliken populations. The functional form of the \(\gamma\) depends on the kind of the DFTB-type method, whereas DFTB2 uses the integral of the two Slater densities and in GFN-xTB the Klopman--Dewar--Sabelli--Ohno~(KDSO) \cite{Klopman1964,Dewar1977,Thiel1992} approximation is used.
The KDSO approximation is also applied in NDDO-type methods for evaluating the two-electron integrals, which will be described in Sec.~\ref{sec:nddo-ewald}.
Apart from different parameter sets for \(\gamma\), the third-order term that describes the change of the chemical hardness
on an atom with \(\Gamma\) is not included in the DFTB2 method.

The electron-electron interactions are divided into three main components: Coulomb repulsion, non-local Fock exchange, and semi-local exchange-correlation.
Both NDDO- and DFTB-type methods include the Coulomb repulsion contribution.
Additionally, NDDO-type methods add non-local Fock exchange, while DFTB-type methods include semi-local exchange-correlation.

A detailed description of semi-empirical NDDO-type and DFTB-type methods can be found in the original publications.\cite{AM1, PM6, DFTB2, GFN1xTB}

\section{Efficient Ewald summation for NDDO-type methods}
\label{sec:nddo-ewald}

The electrostatic interactions present in NDDO-type methods, like AM1 and PM6, are evaluated using atomic point multipoles in CP2K.\cite{cp2k2020}
For this, the two-electron repulsion integral is expressed in an atomic point multipole basis with up to quadrupole moments.
The Coulombic interaction between the multipole moments is approximated using the KDSO screening function~\(\gamma\), given as
\begin{equation}
    \gamma_\text{AB}
    = \frac{s(R_\text{AB})}{R_\text{AB}}
    = \frac1{\bigl(R_{AB}^2 + \left(\rho^\text{A} + \rho^\text{B}\right)\bigr)^{1/2}},
\end{equation}
where \(\rho^\text{A/B}\) are screening parameters for atom \(\text{A/B}\) chosen to recover the correct short-range behavior of the two-electron integral.
Using the KDSO interaction kernel, the two-electron repulsion integral is expressed as
\begin{align}
    \left(\mu_\text{A}\nu_\text{A}|\lambda_\text{B}\sigma_\text{B}\right)
    = \sum_{\substack{\ell=0\\\ell'=0}}^{L_\text{max}}
    \mathbf{M}^{\mu\nu}_{\ell}
    \mathbf{M}^{\lambda\sigma}_{\ell'}
    \bm{\nabla}_\text{A}^{\ell}\bm{\nabla}_\text{B}^{\ell'}
    \frac{s(R_\text{AB})}{R_\text{AB}},
\end{align}
where \(\mathbf{M}^{\mu\nu/\lambda\sigma}_{\ell}\) are the multipole moments of the angular momentum \(\ell\) corresponding to the basis function pair \(\mu\nu\) or \(\lambda\sigma\), \(\bm{\nabla}^{\ell}_\text{A/B}\) is the outer product of the derivative up to rank \(\ell\) with respect to the atomic coordinates of atom \(\text{A/B}\), and the parameter \(L_\text{max}\) controls the highest moment used in the multipole expansion.
For the \textit{sp} basis, an expansion up to \(L_\text{max}=2\) exactly reproduces the two-electron integral within the KDSO approximation.

The KDSO interaction kernel is split into a long~(Eq.~\ref{eq:nddo-lr}) and short range~(Eq.~\ref{eq:nddo-sr}) contribution, i.e.
\begin{align}
    \left(\mu_\text{A}\nu_\text{A}|\lambda_\text{B}\sigma_\text{B}\right)_\text{LR}
    &= \sum_{\substack{\ell=0\\\ell'=0}}^{L_\text{max}}
    \mathbf{M}^{\mu\nu}_{\ell}
    \mathbf{M}^{\lambda\sigma}_{\ell'}
    \bm{\nabla}_\text{A}^{\ell}\bm{\nabla}_\text{B}^{\ell'}
    \frac1{R_\text{AB}}
    \label{eq:nddo-lr}
    \\
    \left(\mu_\text{A}\nu_\text{A}|\lambda_\text{B}\sigma_\text{B}\right)_\text{SR}
    &= \sum_{\substack{\ell=0\\\ell'=0}}^{L_\text{max}}
    \mathbf{M}^{\mu\nu}_{\ell}
    \mathbf{M}^{\lambda\sigma}_{\ell'}
    \bm{\nabla}_\text{A}^{\ell}\bm{\nabla}_\text{B}^{\ell'}
    \frac{s(R_\text{AB})-1}{R_\text{AB}}.
    \label{eq:nddo-sr}
\end{align}
The long range contribution corresponds to the unscreened Coulomb interaction between atomic point multipoles and can be evaluated efficiently using the multipolar Ewald summation.\cite{MultipolarEwald,MultipolarEwaldReciprocalSpace}
For the short range contribution, the leading term decays with \(R_\text{AB}^{-3}\) and can be evaluated either by a dipolar Ewald summation or a single real space cutoff.
The comparison between the real space and dipolar Ewald summation shows only negligible differences, therefore the former approach is preferred due to its simplicity.

\section{\label{sec:comp}COMPUTATIONAL DETAILS}

The CP2K package was used for all MD simulations.\cite{cp2k2020}
Specifically, the NDDO-type (AM1,\cite{AM1} \AM1W{},\cite{AM1W} PM6,\cite{PM6} and \PM6fm{}\cite{PM6fm}) and
DFTB-type (DFTB2\cite{DFTB2} and DFTB-iBi\cite{DFTB2iBi}) models were considered, as well as the recently devised \GFNxTB{} method.\cite{GFN1xTB} Throughout, periodic boundary conditions (PBCs) are employed using the Ewald summation method.\cite{Ewald1921} Therein, the charge distribution is split into short- and long-range terms by adding and subtract a Gaussian distribution with $\alpha = 1/(\sqrt{2}\sigma)= 0.35$~\AA{}$^{-1}$, where $\sigma$ is the standard deviations of the employed Gaussian distribution. 
The multipolar Ewald summation was used in the periodic NDDO-type calculations,
\cite{MultipolarEwald, MultipolarEwaldReciprocalSpace}
because the electrostatic energy is comprised of multipole-multipole interactions.
By contrast, since merely monopole terms are entailed in the DFTB-type methods
for electron density fluctuations,
the more efficient smooth particle mesh Ewald method\cite{SPME}
was applied to the periodic DFTB2, \DFTB2iBi{}, and \GFNxTB{} simulations. 
After experimenting with some trial setups, we selected %$\alpha = 0.35$~\AA{}$^{-1}$ for the background neutralizing charge distribution and
$2$ grid points per \AA{} for the long-range summation in the reciprocal space.
We would like to point out that the treatment of PBCs does not
form a computational bottleneck in our MD simulations, which is dominated by solving the generalized eigenvalue problem.\cite{schade2022towards, schade2023breaking} %when the relevant settings have been wisely selected.
In our present study, the computational efficiency of the SQC methods was observed to be such that
% some hardware details are intentionally avoided here. No advertisement for Intel.
one MD step usually took less than $1$~second on a single computing node with
two Intel Xeon CPUs at 2.4~GHz.

Car-Parrinello-like Born-Oppenheimer MD simulations were performed with 128 light water molecules
in a periodic box of length $L = 15.6404$~\AA{}, resulting in a density of $1.0$~g/cm$^3$,
in the canonical (NVT) ensemble.\cite{Kuehne2007PRL,Kuehne2014WIRE}
The temperature was held constant at $300$~K using the method of Nosé and Hoover,
\cite{Nose1984, Hoover1985} with chain thermostats\cite{MassThermo} applied to
all degrees of freedom (so-called ``massive'' thermostatting).
A time step of $0.5$~fs and a convergence criterion of
$10^{-7}$~a.u.\ for the wave function were employed throughout all MD simulations.
The starting configuration was taken from a well-equilibrated simulation
in our previous study.%
\cite{Kuehne2013NC}
The system was, nevertheless, further equilibrated for $25$~ps
by using the respective SQC methods.
We also confirmed the equilibration by checking
the variations of temperature and potential energy during the MD simulation.
It was then followed by production runs for $125$~ps each.

For comparison, we have also performed DFT-based second-generation Car-Parrinello AIMD simulations at the \BLYPD3{}/TZV2P level of theory of the same system. The \BLYPD3{} XC functional has been shown to be a robust computational method for liquid water properties at ambient conditions in preceding studies,\cite{BLYPDisGood, Kuehne2011JPCL, Kuehne2013NC, Kuehne2014JACS, Elgabarty2015NC,Elgabarty_TDK_SA2020} and we use it here to benchmark the aforementioned SQC methods for some liquid water properties, which are not experimentally available.

\section{\label{sec:resndis}RESULTS AND DISCUSSION}

The accuracy of the employed SQC methods is assessed by means of various static and dynamical properties. Whereas the former are simple Boltzmann-weighted ensemble averages, which are obtained as time averages along our MD trajectories, the latter are typically computed via appropriate time-correlation functions. 

\subsection{\label{sec:static}Static properties}

\subsubsection{\label{sec:radial}Pair distribution function}

\begin{figure}
\includegraphics[width=0.42\textwidth]{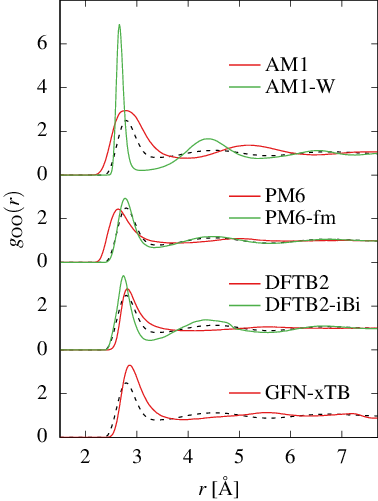}%
\caption{\label{fig:rdfOO}Oxygen-oxygen radial distribution functions \gOO{} calculated with different SQC methods. The reference \gOO{} (dashed lines) is from neutron total scattering experiments.\cite{Soper2013}}%
\end{figure}

The intermolecular oxygen-oxygen radial distribution functions (RDFs)\cite{rohrig2013optimal} \gOO{} for liquid water 
simulated by using various SQC methods are shown in Fig.~\ref{fig:rdfOO}.
Detailed quantitative features obtained from these \gOO{} are summarized in Table~\ref{SItab:rdfOO} in the supplementary information.
It is apparent that the original parametrizations of all considered SQC methods (see red lines in Fig.~\ref{fig:rdfOO})
fail to describe the first solvation shell; either the position or the width of the first peak are strongly deviating from the experimental reference (dashed lines), with a rather flat first minimum that is shifted towards
larger intermolecular \ce{O-O} distances, and a grossly overestimated coordination number (see $N_\text{c}$ in Table~\ref{tab:LiqWatProp}).%
\begin{table}
\caption{\label{tab:LiqWatProp}Properties of liquid water at ambient conditions
  obtained from our MD simulations with different SQC methods.
  \DHvap{} is the heat of vaporization (in kcal/mol).
  $D_\text{PBC}$ is the translational diffusion coefficient (in $10^{-5}$~cm$^{2}$/s) with periodic boundary conditions
  and $N_\text{c}$ is the coordination number calculated by
  integrating \gOO{} till the first minimum.}
\begin{ruledtabular}
\begin{tabular}{cddd}
  Method                           &
  \multicolumn{1}{c}{\DHvap{}}     &
  \multicolumn{1}{c}{$D_\text{PBC}$}          &
  \multicolumn{1}{c}{$N_\text{c}$} \\
  \hline
  AM1         &  8.20 &  2.917 & 10.3 \\
  \AM1W{}     & 17.62 &  0.010 &  3.8 \\
  PM6         &  6.83 & 10.899 &  6.6 \\
  \PM6fm{}    &  9.32 &  1.540 &  4.5 \\
  DFTB2       &  3.98 &  9.331 &  8.2 \\
  \DFTB2iBi{} &  5.38 &  1.032 &  4.0 \\
  \GFNxTB{}   & 11.44 &  4.497 &  8.9 \\
  Expt.       & 10.50 &  2.395 &  4.7 \\
\end{tabular}
\end{ruledtabular}
\end{table}
The second solvation shell is severely underestimated and shifted by $0.5$ to $1.0$~\AA{} towards larger distances, whereas the second minimum and the third peak are barely, if at all, observable.

\begin{figure}
\includegraphics[width=0.42\textwidth]{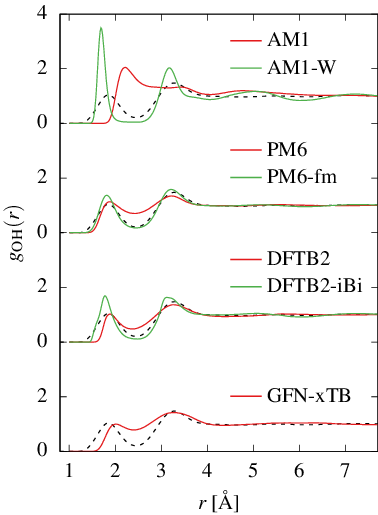}%
\caption{\label{fig:rdfOH}Oxygen-hydrogen radial distribution functions 
   \gOH{}, as obtained by different SQC methods.
  The reference \gOH{} (dashed lines) is from
  neutron total scattering experiments.}%
\end{figure}

% gOH(r)
Fig.~\ref{fig:rdfOH}%
shows the intermolecular \gOH{} is a function of the employed SQC method.
The most noticeable failure of AM1 is the absence of the first minimum,
with a shifted first peak that overlaps the second peak to a large extent.
The other original SQC methods, i.e. PM6, DFTB2, and \GFNxTB{},
predict only a very shallow first minimum.
All these results show that using the original parametrizations, these SQC methods are not able to
describe the radial structure of liquid water properly.\cite{RuizLopez2005,SCPNDDO4water,DFTB2gammaHB4Water,Voth2010,Voth2013,Voth2014,QC2011}
At last, it is interesting to note that DFTB2 and \GFNxTB{} produce rather similar RDFs, which may be attributed to similarities that both methods share. 

% specific reparametrized SQC methods
The performance of the specifically reparametrized
SQC methods, i.e. \AM1W{}, \PM6fm{}, and \DFTB2iBi{} are rather different.
Their \gOO{} and \gOH{} are plotted as
solid green lines in Figs.~\ref{fig:rdfOO} and \ref{fig:rdfOH}, respectively,
and again also compared with the experimental reference.
% AM1-W
%The \AM1W{} parameters were optimized for the structures and energies of small water clusters (up to trimer at most) calculated at the DFT level.\cite{AM1W}
The first peaks in all RDFs of \AM1W{} (see also \gHH{} in Fig.~\ref{SIfig:rdfHH})
are unreasonably high and narrow compared with the experimental curves.
The second peaks are also more pronounced than those in experiment.
Moreover, the presence of significant density depletion in the interstitial region
between the first two solvation shells reflects a highly ordered short-range structure.
Analyzing the MD trajectory produced by the \AM1W{} model,
an amorphous structure consisting of rigid water molecules becomes apparent.
% DFTB2-iBi
Compared to experiment, the RDFs produced by \DFTB2iBi{} indicate
a somewhat overstructured liquid water, and the
two peaks within \gOH{} that are almost equal in height. Interestingly, 
these observations are consistent with previous DFT-based AIMD simulation using the PBE XC functional,\cite{Kuehne2009JCTC}
which was actually used as reference while parametrizing \DFTB2iBi{}.\cite{DFTB2iBi}
% PM6-fm
Overall, the RDFs obtained with \PM6fm{} exhibit the best agreement with experiment. Even tough the first peaks of \gOO{}, \gOH{} and \gHH{} are slightly higher than those observed in experiment, they are expected to be reduced
when nuclear quantum effects are taken into account,%
\cite{Markland2016, Kuehne2015MP} %, Kuehne2019CPC}
thereby further increasing the agreement with experiment. 
% remark
%The above results also demonstrate that the use of different reference systems of water in a particular reparametrization of an SQC method does have a significant impact on the accuracy for describing liquid water properties at ambient conditions.

\subsubsection{\label{sec:angular}Angular structure}

\begin{figure}
\centering
\begin{tabular}{@{}%
  p{0.22\linewidth}@{\quad}p{0.22\linewidth}@{\quad}%
  p{0.22\linewidth}@{\quad}p{0.22\linewidth}@{}}
  \subfig[width=\linewidth]{AM1}{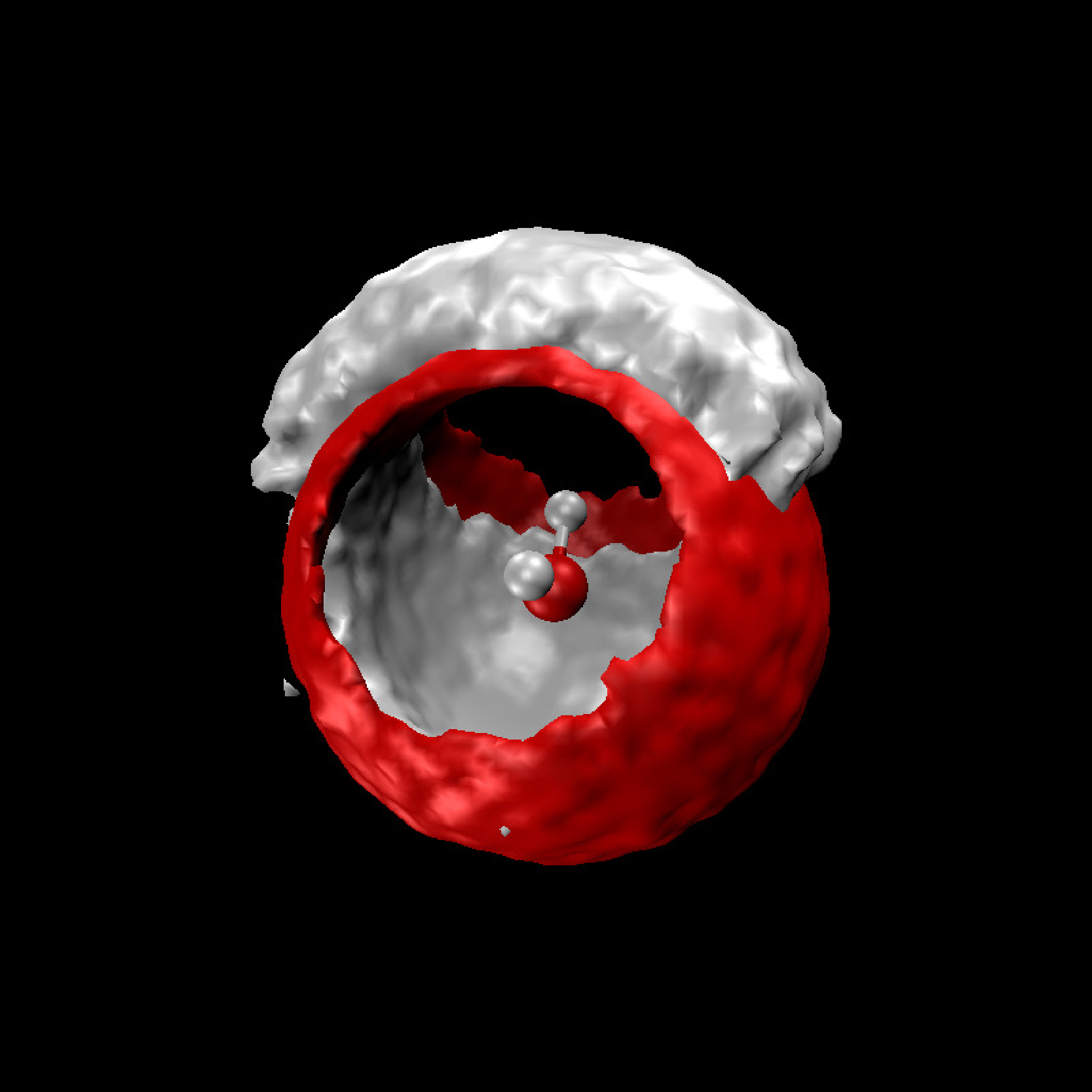}              &
  \subfig[width=\linewidth]{PM6}{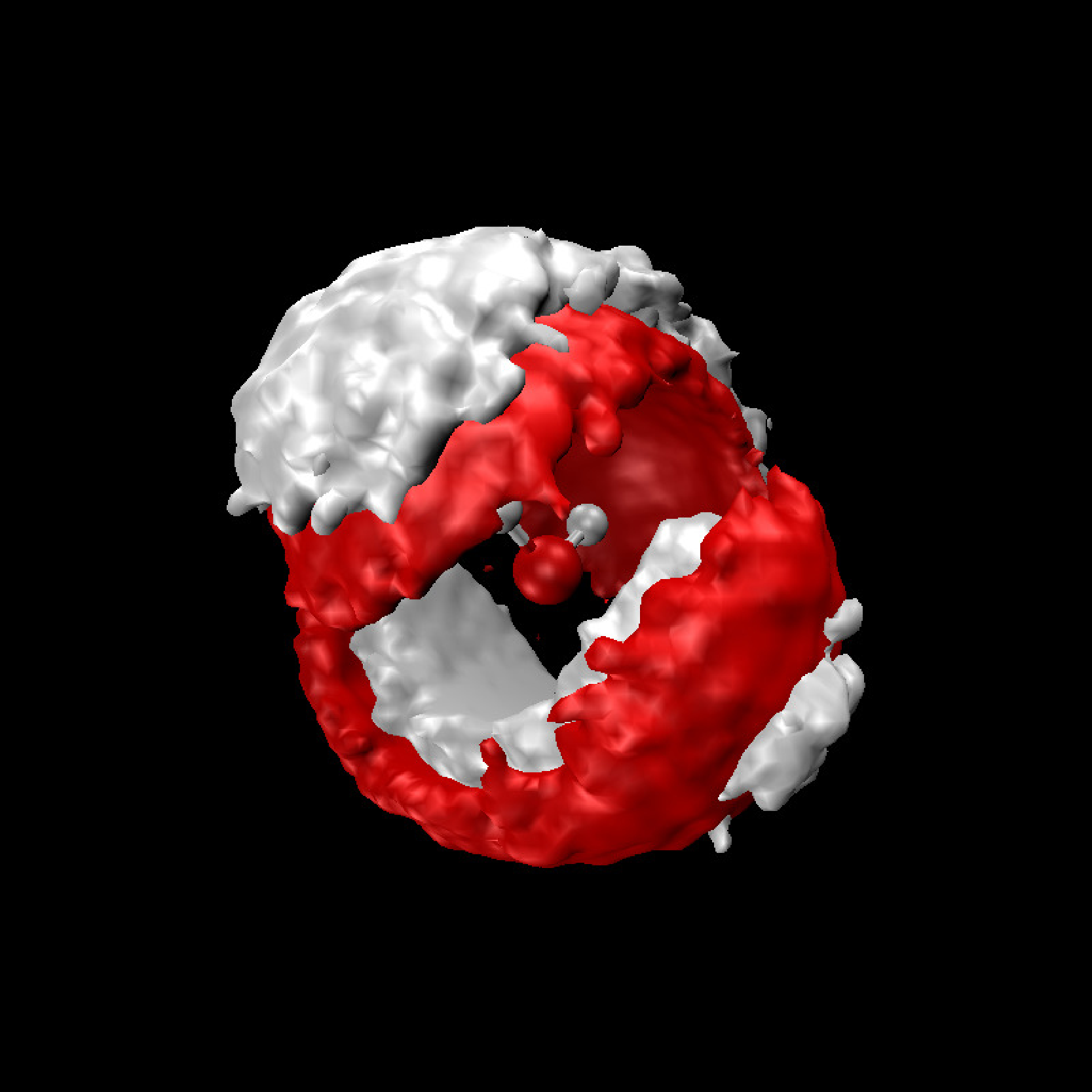}              &
  \subfig[width=\linewidth]{DFTB2}{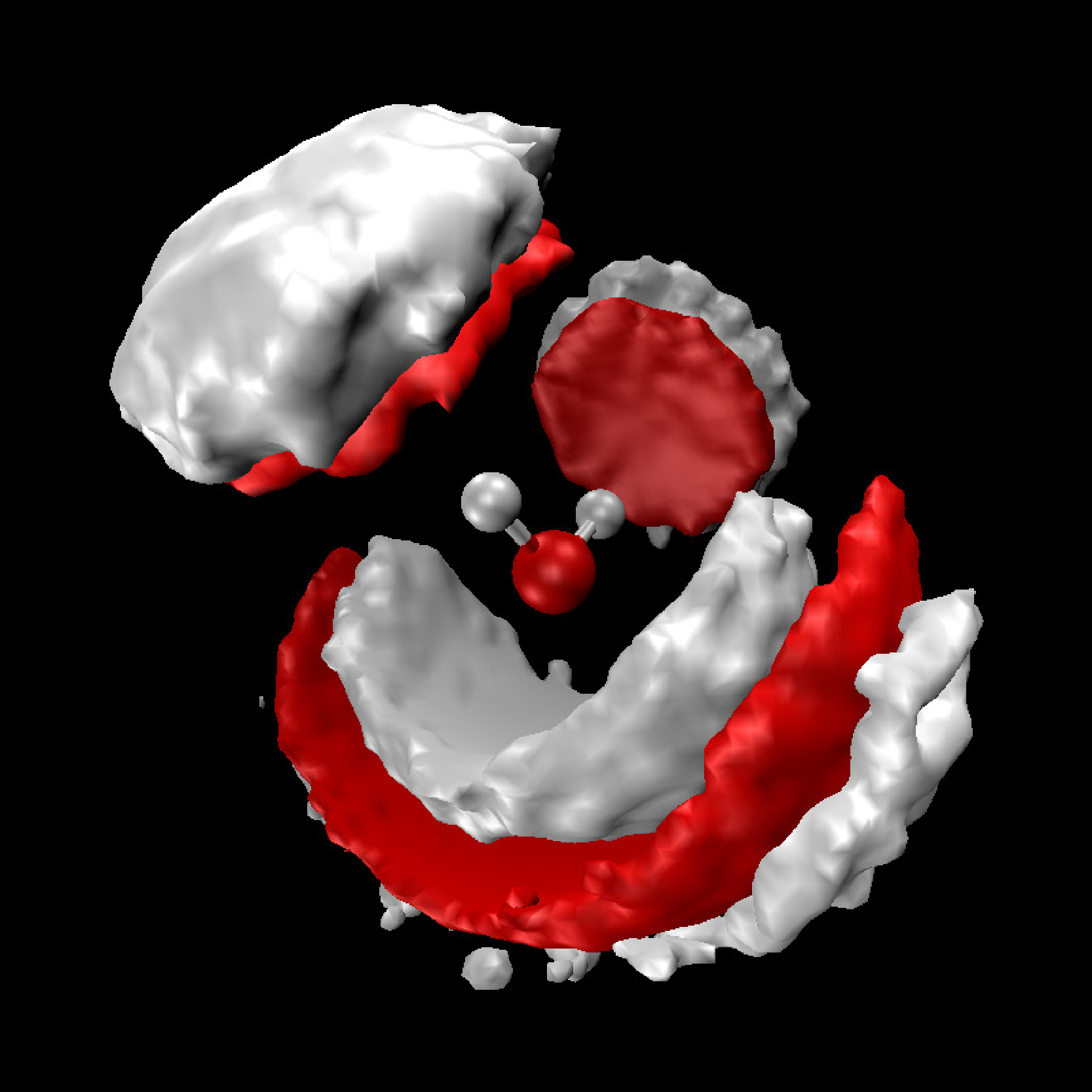}          &
  \subfig[width=\linewidth]{\GFNxTB{}}{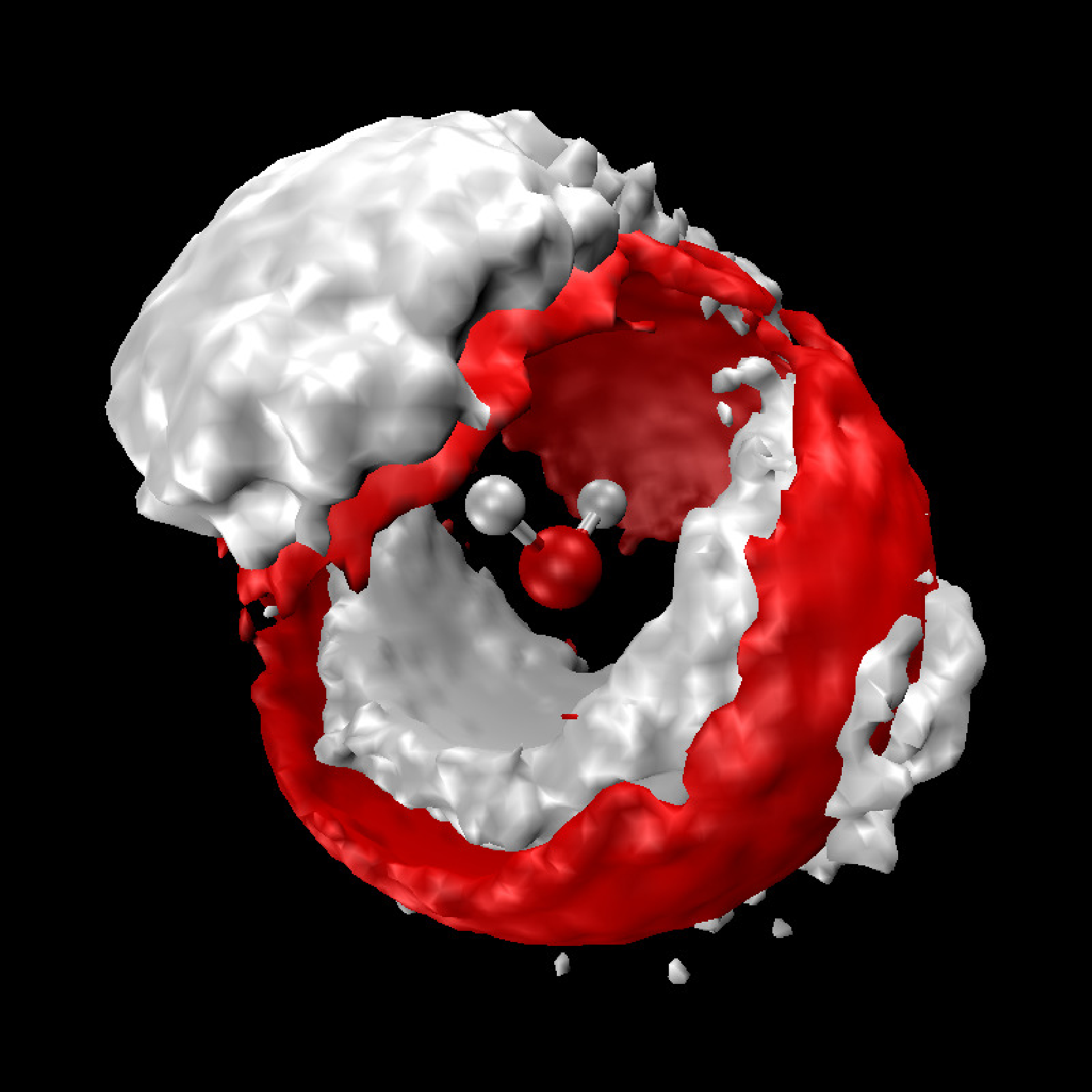}     \\[1sp]
  \subfig[width=\linewidth]{\AM1W{}}{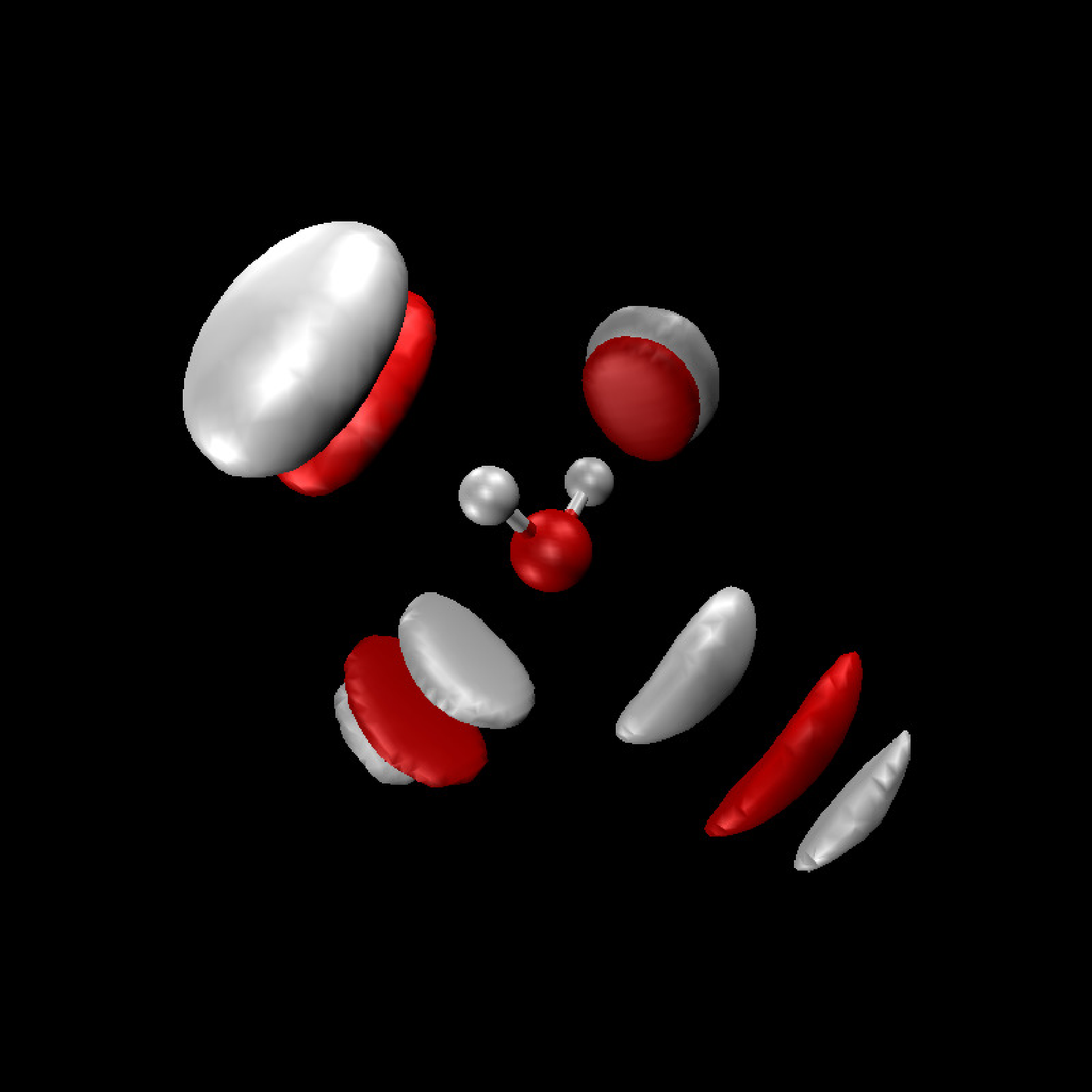}         &
  \subfig[width=\linewidth]{\PM6fm{}}{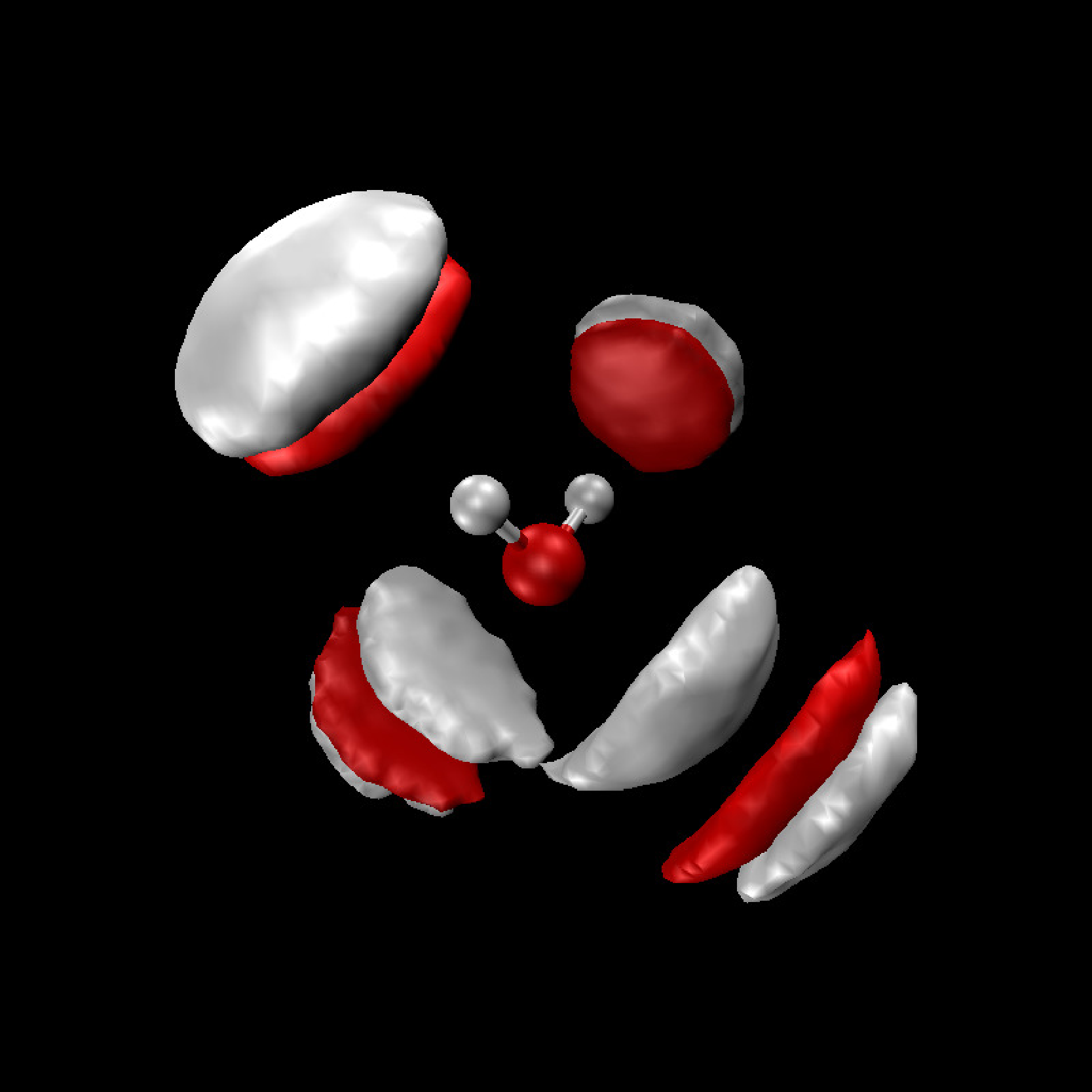}       &
  \subfig[width=\linewidth]{\DFTB2iBi{}}{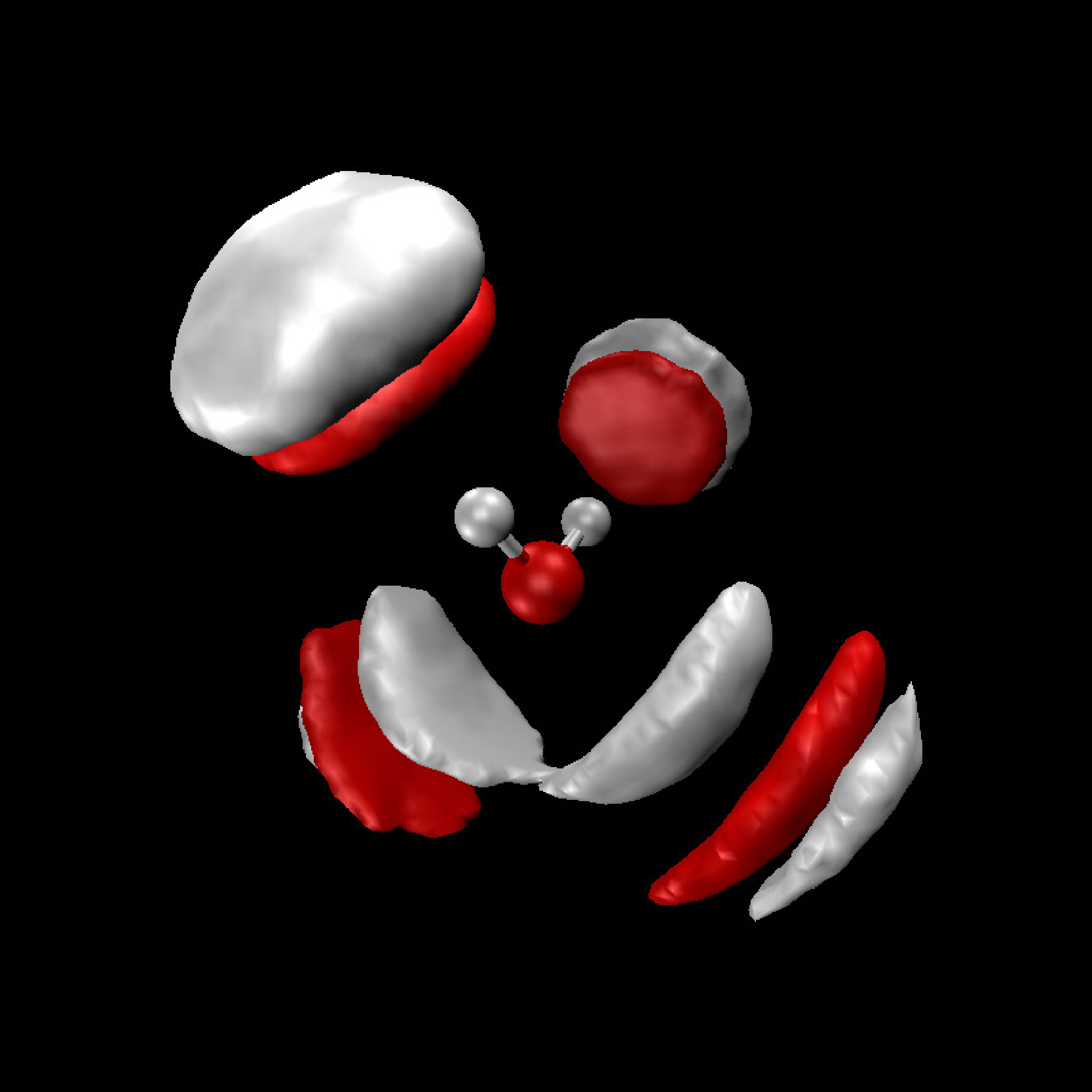} &
  \subfig[width=\linewidth]{\BLYPD3{} DFT}{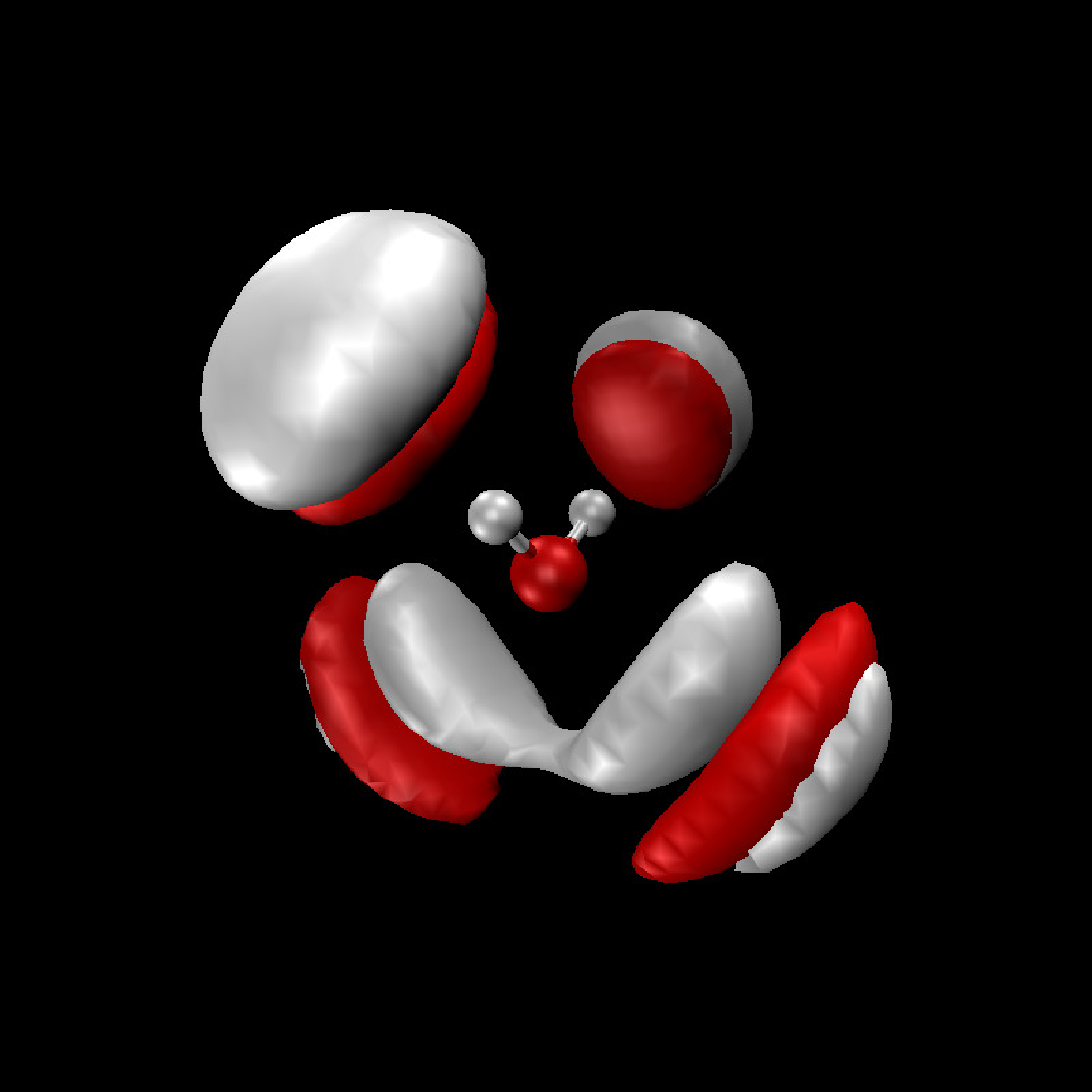}
\end{tabular}
\caption{\label{fig:SDFs}Spatial distribution functions of the four nearest neighbors
  surrounding a central water molecule calculated with different SQC methods and DFT with the \BLYPD3{} XC functional.
  Oxygen and hydrogen atoms are colored in red and white, respectively.
  The distributions were obtained by using ANGULA~\cite{ANGULA} program and plotted with VMD~\cite{VMD} at an isosurface density fraction of $0.85$.}%
\end{figure}

\begin{figure}
\includegraphics[width=0.45\textwidth]{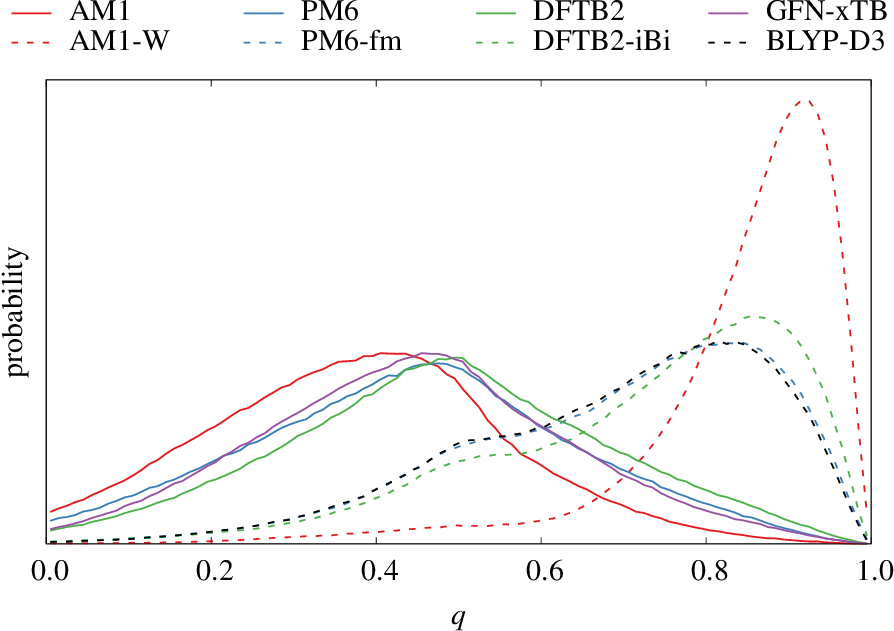}%
\caption{\label{fig:orderQ}Probability distribution of
  the orientational order parameter $q$ for liquid water at ambient conditions, as obtained by different SQC methods and DFT using the \BLYPD3{} XC functional.}%
\end{figure}

To assess the local angular structure of simulated liquid water 
when using different SQC methods, the spatial distribution functions (SDFs) of
the first four neighboring water molecules around a central water and
the probability distribution for the orientational order parameter $q$ are shown in
Figs.~\ref{fig:SDFs} and \ref{fig:orderQ}, respectively. 
These plots, first proposed for water by Svishchev and Kusalik,~\cite{FirstSDFKusalik1993} offer a view into the local angular correlation structure of liquids. Soper~\cite{SDFSoper1994} has also derived it from neutron scattering experiments via reconstruction of the orientational correlation function consistent with the measured data. % The \BLYPD3{} DFT results are also shown for comparison.


% SDFs
The original parametrizations of the here considered SQC methods (first row in Fig.~\ref{fig:SDFs}), especially the NDDO-type approaches, predict loosely bound water as H-bond acceptor (and also as H-bond donor) to the central water molecule,
which suggests an overall lack of angular structure. On the contrary, all reparametrized SQC methods (as well as \BLYPD3{} DFT) leads to a much more defined orientation for the two H-bond acceptor water molecules (see second row in Fig.~\ref{fig:SDFs}). The H-bond donors in \AM1W{}, however, tend to have an extremely constrained coordination pattern.
%which resembles the amorphous ice structure again.
At variance, both DFTB-iBi and \PM6fm{} schemes are in much closer agreement with
the arrangement at the \BLYPD3{} DFT level, in which the H-bond donor molecules are
slightly more delocalized than the H-bond acceptors~\cite{Kuehne2013NC}.

% q
The orientational order parameter $q$, which is depicted in Fig.~\ref{fig:orderQ}, characterizes the local angular structure of liquid water as deviating from the ideal tetrahedral angles of ice $I_h$. It is defined as
\begin{equation*}
  q = 1 - \frac{3}{8}\sum_{i=1}^{3}\sum_{j=i+1}^{4}
  \left( \cos\psi_{ij} + \frac{1}{3} \right)^{2},
\end{equation*}
where $\psi_{ij}$ is the angle between the two vectors pointing from
the central oxygen to the nearest-neighbor oxygen atoms, which are labeled $i$ and $j$.%
\cite{Debenedetti2001} %, Hardwick1998, Laage2015}
The two extremes of $q$ are 0 for an ideal gas and 1 for a perfect tetrahedron, respectively.
% original SQC methods
As can be seen in Fig.~\ref{fig:orderQ}, the probability distributions of $q$
for the original parametrizations of all SQC methods are centered around $0.5$,
revealing a strongly perturbed tetrahedral structure, which is consistent with
the associated SDFs shown in the first row of Fig.~\ref{fig:SDFs}.
The corresponding distribution of the AM1 approach bends toward even smaller values. By contrast, the reparametrized SQC method \AM1W{} possesses a notably high distribution of $q$ around $0.9$, in agreement with its most rigid angular structure. The \PM6fm{} result, however, is indistinguishable from the curve of our \BLYPD3{} DFT simulations, whereas \DFTB2iBi{} favors a slightly more ordered tetrahedral structure.

\subsubsection{\label{sec:HBNetwork}H-bond network}

\begin{figure*}
\centering
\begin{tabular}{@{}%
  p{0.21\linewidth}@{\qquad}p{0.21\linewidth}@{\qquad}%
  p{0.21\linewidth}@{\qquad}p{0.21\linewidth}@{}}
  \subfigWRBeta[width=\linewidth]{AM1}{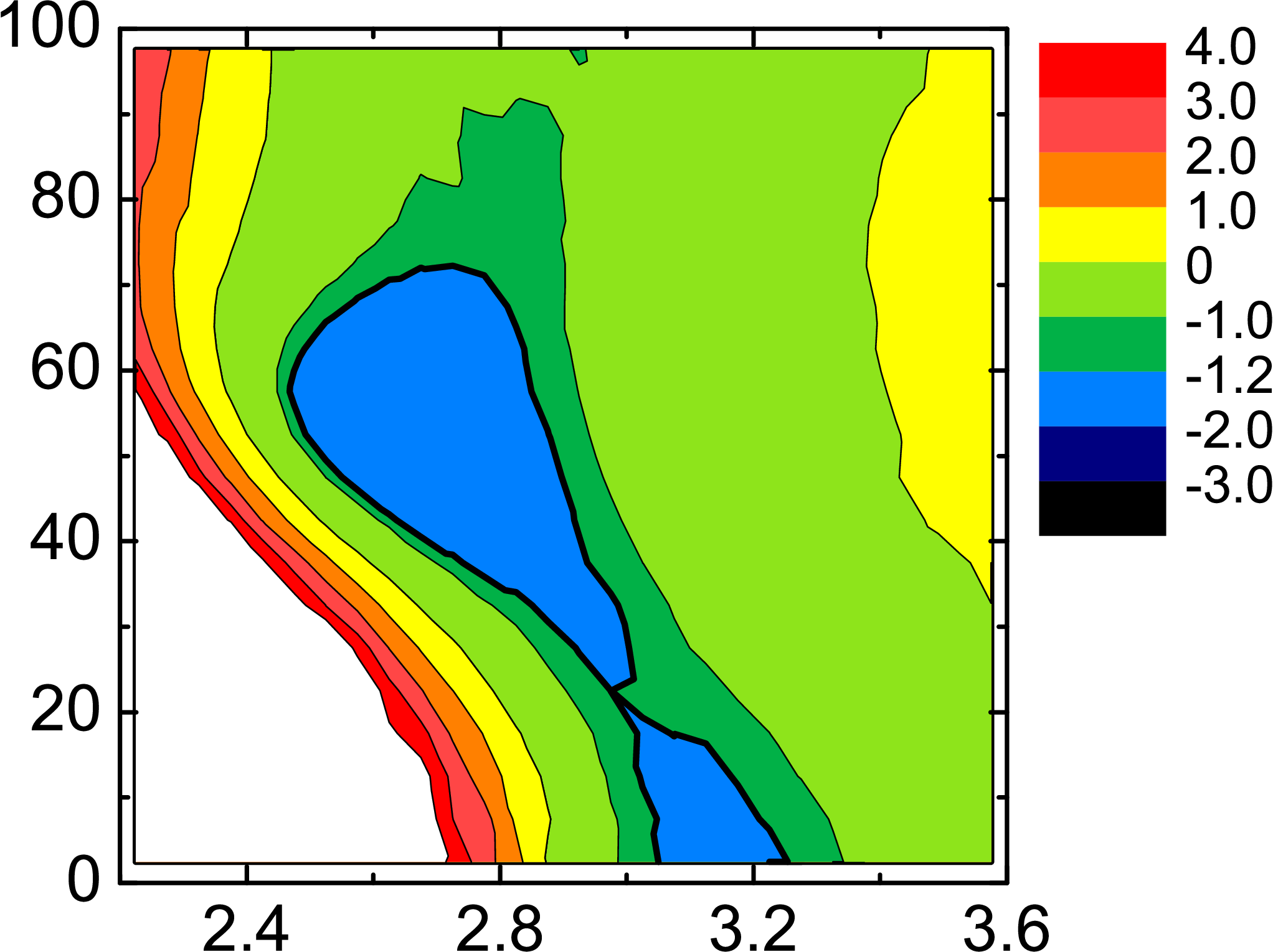}              &
  \subfigWRBeta[width=\linewidth]{PM6}{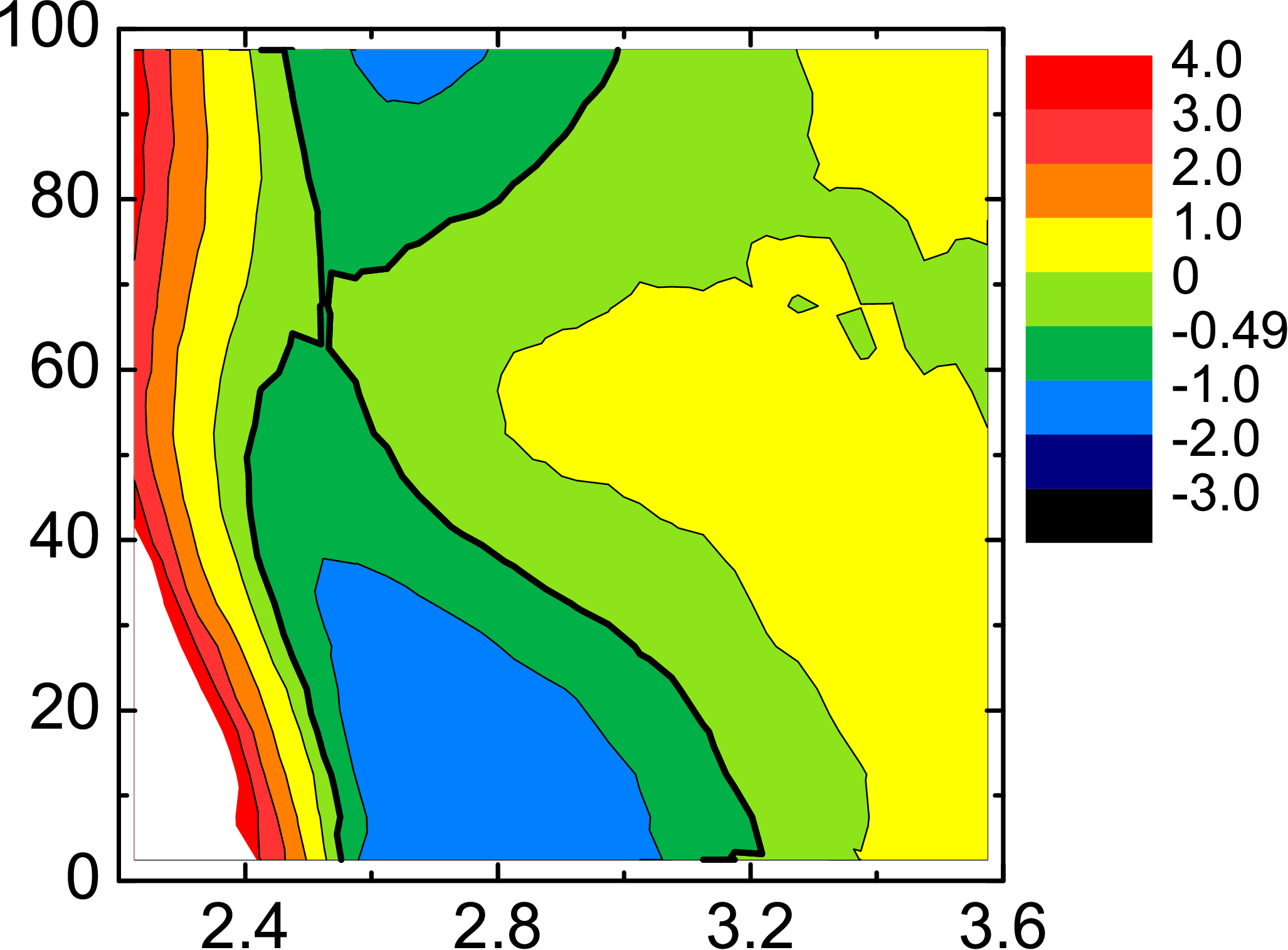}              &
  \subfigWRBeta[width=\linewidth]{DFTB2}{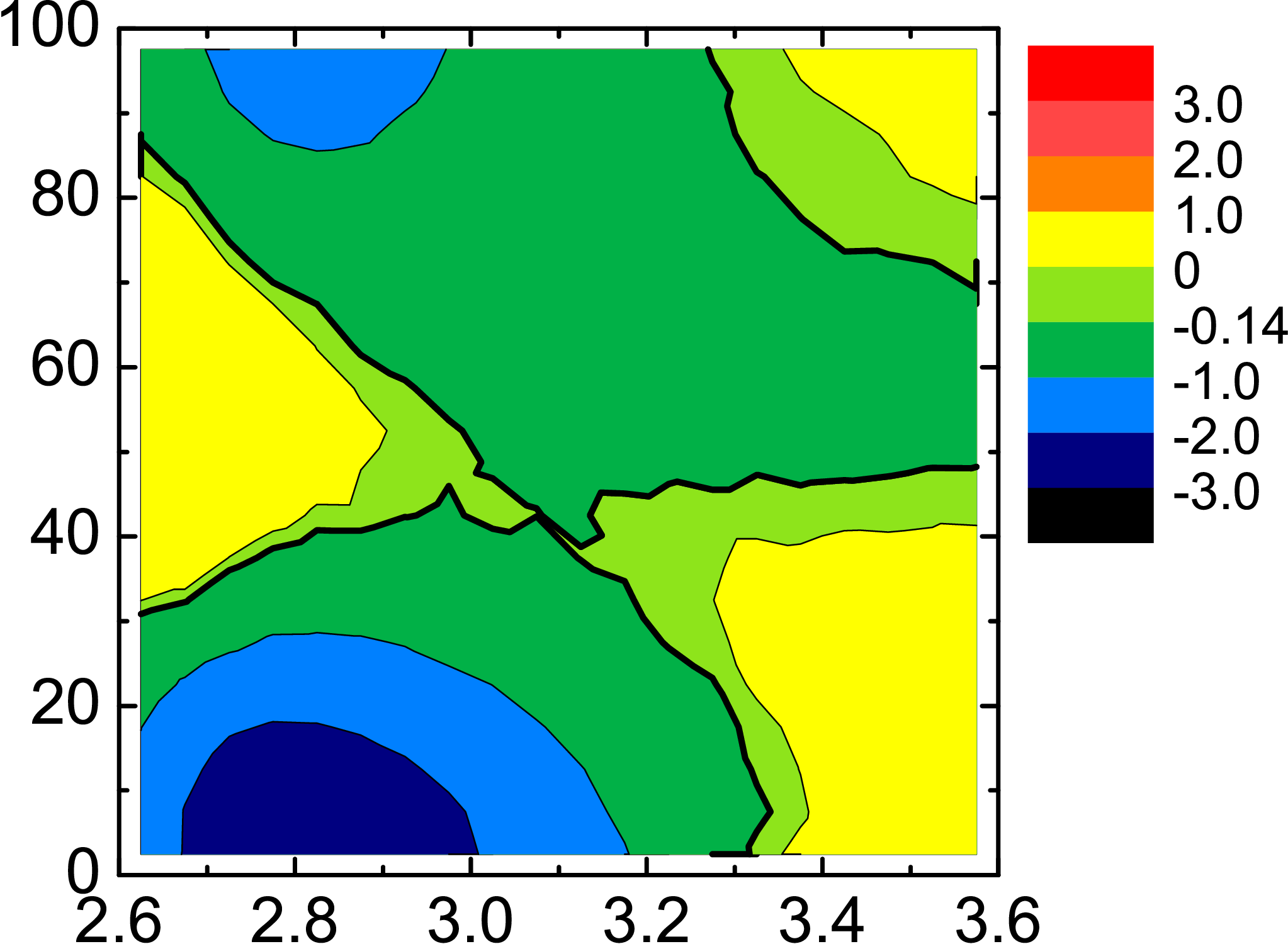}          &
  \subfigWRBeta[width=\linewidth]{\GFNxTB{}}{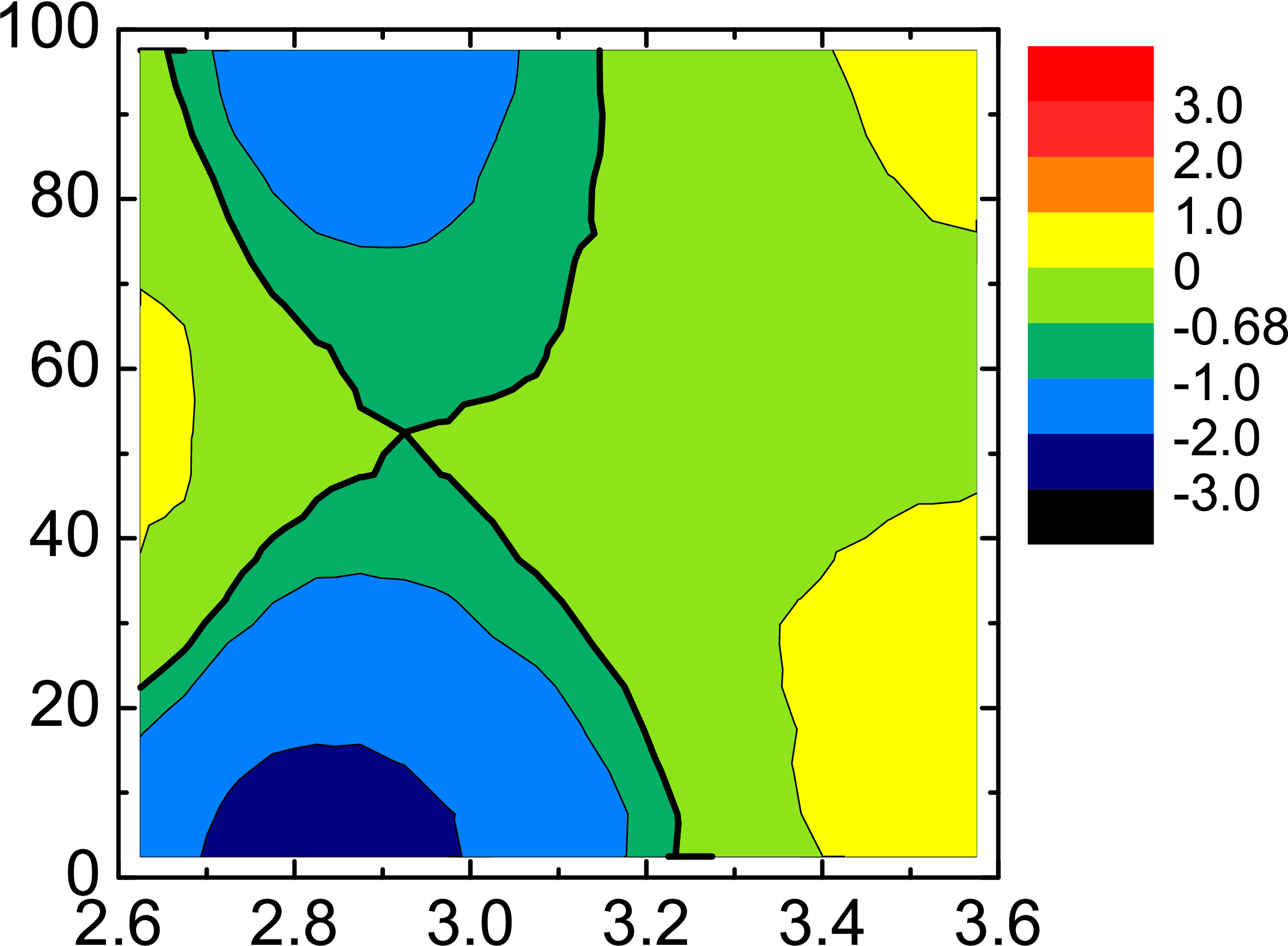}     \\[1sp]
  \subfigWRBeta[width=\linewidth]{\AM1W{}}{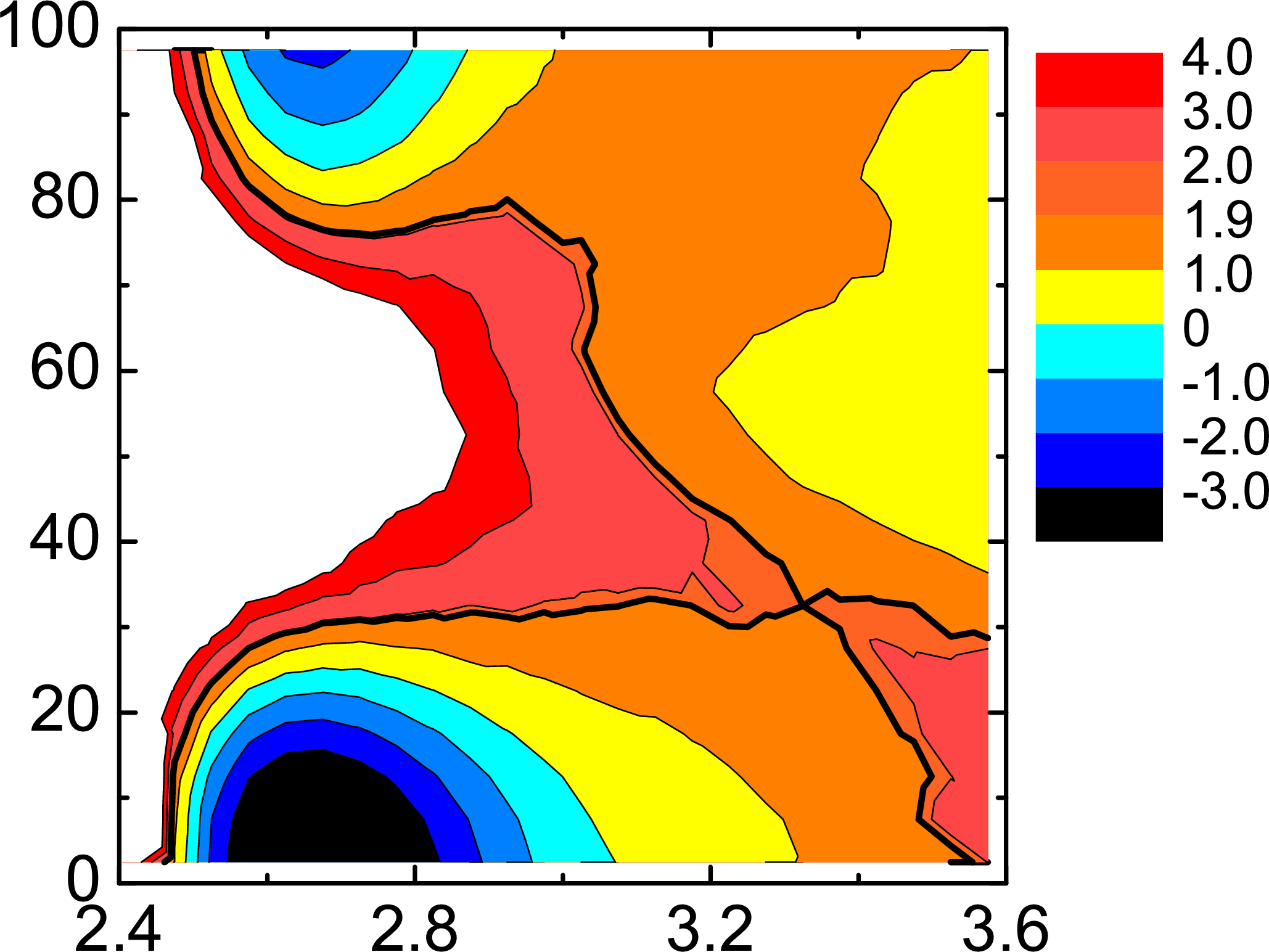}         &
  \subfigWRBeta[width=\linewidth]{\PM6fm{}}{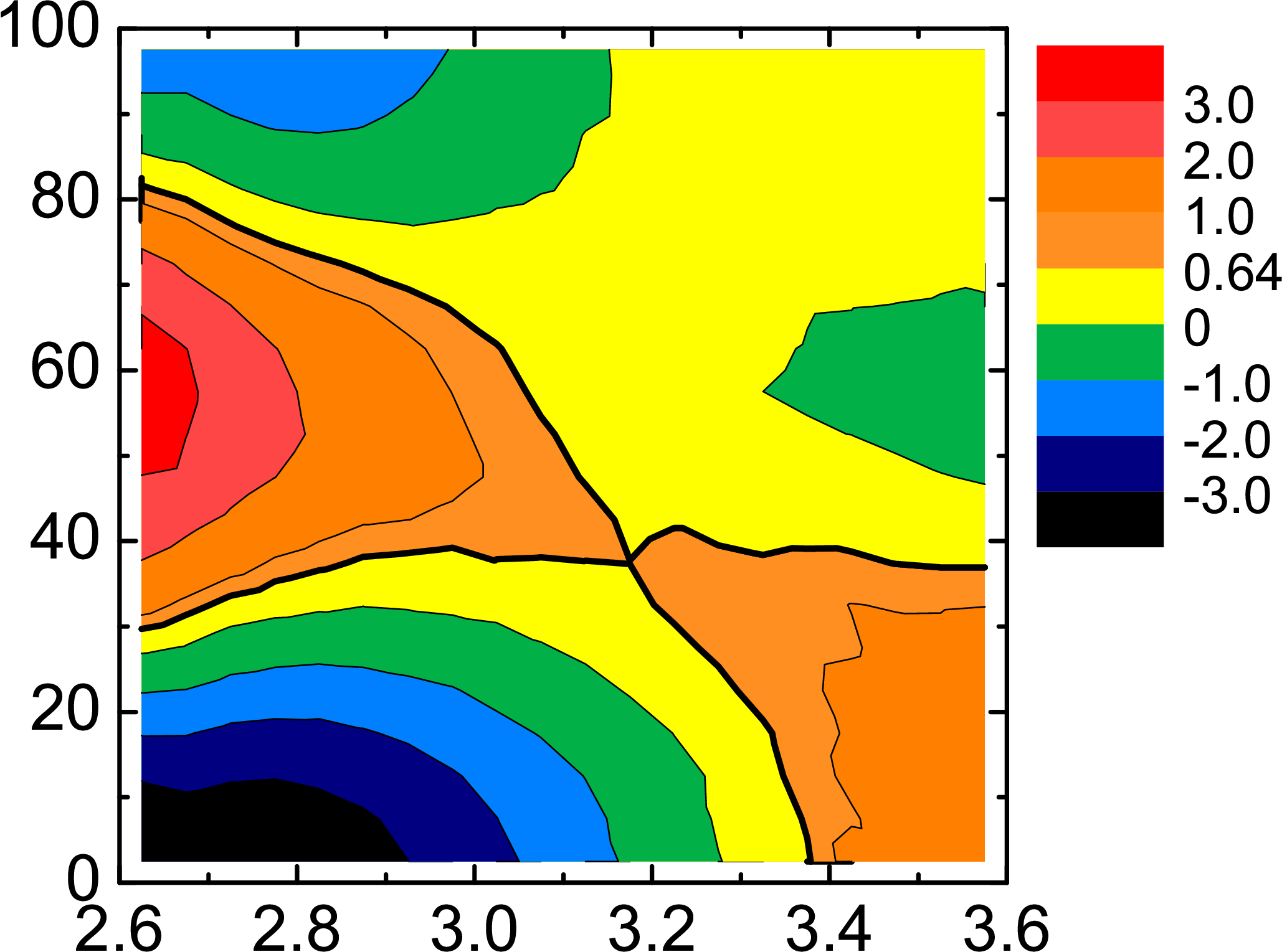}       &
  \subfigWRBeta[width=\linewidth]{\DFTB2iBi{}}{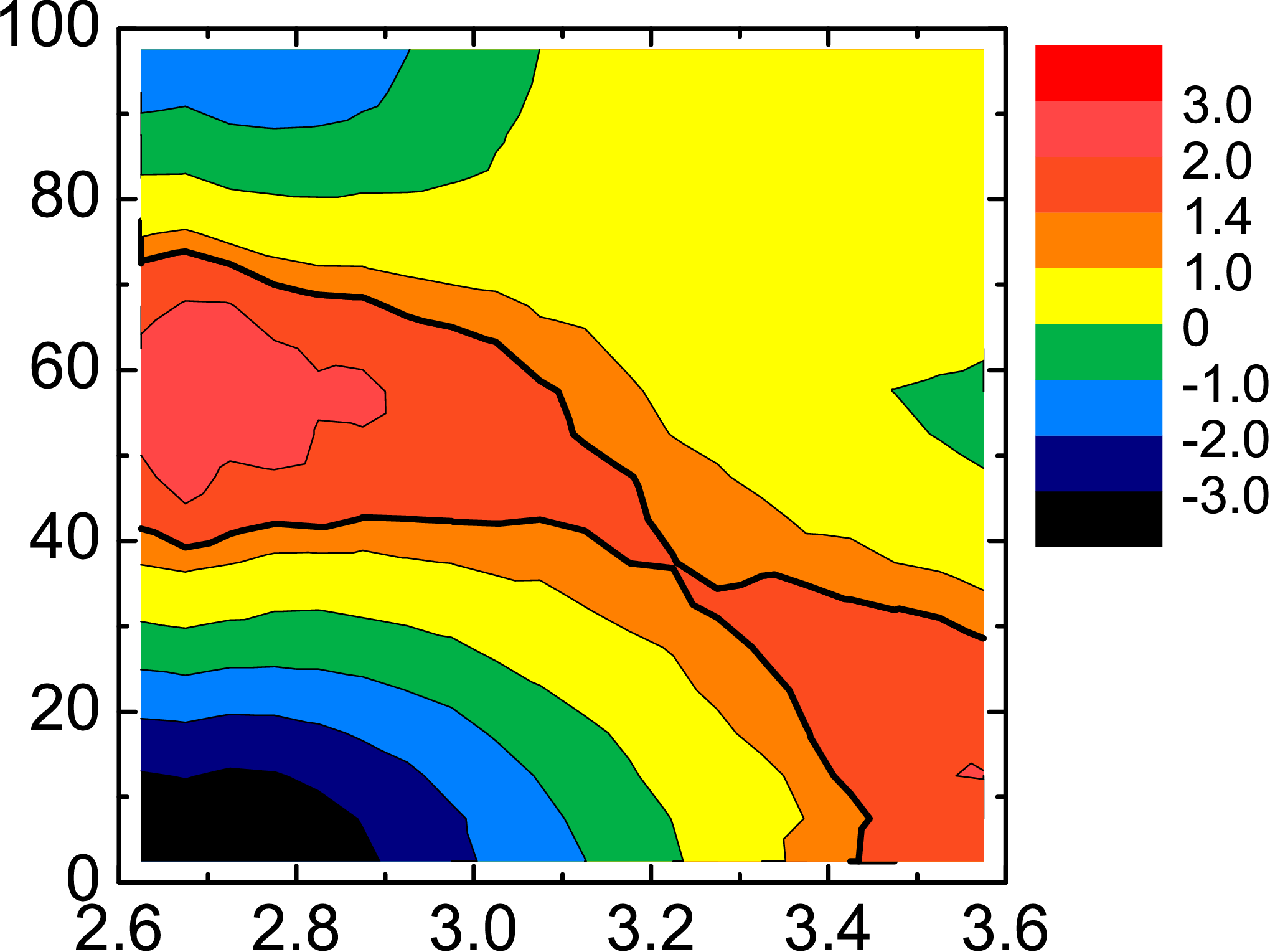} &
  \subfigWRBeta[width=\linewidth]{\BLYPD3{} DFT}{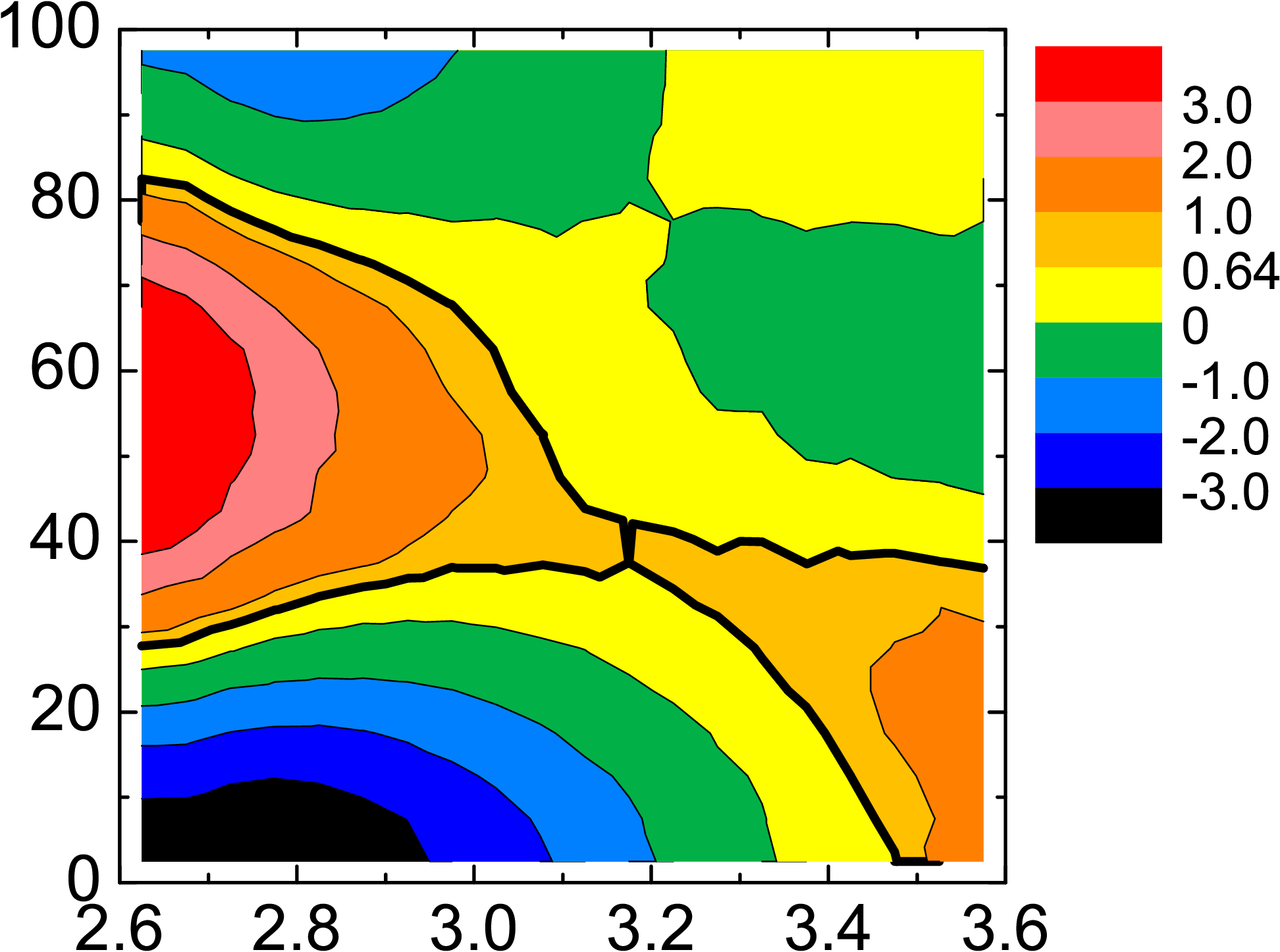}
\end{tabular}
\caption{\label{fig:Rbeta}Contour plot of \WRbeta{} potential of mean force of an
  H-bond in liquid water at ambient conditions as 
  calculated with different SQC methods and \BLYPD3{} DFT.
  The contours are all in units of $k_BT$.}%
\end{figure*}

The H-bond plays a determining role in the structure and dynamics of liquid water.\cite{Tanaka93HB, Henao2016, Kuehne2013NC, Elgabarty_TDK_PCCP2020}
This has been studied here by using the joint radial-angular \Rbeta{} distribution\cite{RBeta} and constructing its corresponding
potential of mean force as defined by Kumar et al., \cite{HBondDef}
i.e. 
\begin{equation*}
  W(R,\beta) = -k T\ln g(R,\beta), 
\end{equation*}
where $R$ is the oxygen-oxygen distance and $\beta$ is the \ce{O-H\bond{...}O} angle
(see Fig.~\ref{SIfig:HBDef}).
The results are plotted in Fig.~\ref{fig:Rbeta}%
and show a minimum basin corresponding to H-bonded configurations.
This minimum basin is separated from the non-H-bonded surface by the equipotential line
where a saddle point is found -- as an example, this saddle point can be seen at $3.2$~\AA{} and $40$ degrees for the \PM6fm{} method.

On a qualitative level, the contour plots show similarities between the NDDO-type methods AM1 and PM6, as well as between the DFTB-type methods DFTB2 and \GFNxTB{}, respectively. Also, the reparametrized SQC methods reflect a similar H-bond geometry. Quantitative results can be extracted by doing an analysis of the H-bonded molecules contained within the minimum basin. The number of H-bonds, fraction of different H-bond donors, and free-energy barrier of H-bond breaking $\Delta W$ are presented in Table~\ref{tab:HBond}. %
\begin{table}
\caption{\label{tab:HBond}Average number of H-bonds per water molecule (\nHB{}),
  populations of double-donor (\fDD{}), single-donor (\fSD{}),
  and non-donor (\fND{}) configurations of water molecules,
  and the free-energy barrier of breaking an H-bond ($\Delta W$, in $k_BT$) for
  the \Rbeta{}-definition of H-bonding in liquid water.}
\begin{ruledtabular}
\begin{tabular}{cccccc}
  % {R, β}-definition
  Method     &
  \nHB{}     &
  \fDD{}     &
  \fSD{}     &
  \fND{}     &
  $\Delta W$ \\
  \hline
  AM1         & 3.46 & 0.48 & 0.43 & 0.09 & 0.73 \\
  \AM1W{}     & 3.88 & 0.94 & 0.06 & 0.00 & 7.26 \\
  PM6         & 3.00 & 0.46 & 0.44 & 0.10 & 1.36 \\
  \PM6fm{}    & 3.49 & 0.74 & 0.24 & 0.02 & 4.60 \\
  DFTB2       & 2.97 & 0.52 & 0.41 & 0.07 & 2.78 \\
  \DFTB2iBi{} & 3.36 & 0.69 & 0.28 & 0.03 & 5.42 \\
  \GFNxTB{}   & 3.17 & 0.54 & 0.39 & 0.07 & 1.81 \\
  \BLYPD3{} DFT  & 3.58 & 0.78 & 0.20 & 0.02 & 4.58 \\
\end{tabular}
\end{ruledtabular}
\end{table}%
The values for $\Delta W$ can be compared to the classical water pair potential SPC/E used in the paper of Kumar et al.~\cite{HBondDef} of about $4.2k_BT$; DFT with the \BLYPD3{} functional and \PM6fm{} are both in close agreement with that value, whereas \DFTB2iBi{} has a higher H-bond breaking barrier. This is in agreement with the structural SDF and tetrahedrality parameter $q$, where \DFTB2iBi{} is overstructured in $q$ compared to \PM6fm{} and \BLYPD3{} DFT, respectively. The original NDDO-type methods show a value that is somewhat too low, which means that the H-bond energy in liquid water is not represented well in these methods. Even though, this is improved in the original DFTB-type methods, it is still low compared to reference \BLYPD3{} DFT calculations. The number of H-bonds and fraction of double and single donors, as obtained by the modified \PM6fm{} and \DFTB2iBi{} methods are also in better agreement with the \BLYPD3{} DFT reference values of around $3.4$~\nHB{} and $f_\text{DD}$ of about $0.7$, respectively. Finally, \AM1W{} entails an amorphous ice structure with an almost perfect tetrahedral structure and high H-bond breaking free-energy penalty.

\subsubsection{\label{sec:Hvap}Heat of vaporization}

Out of the enthalpic properties of liquid water, we elect to compute the heat of vaporization, a quantity which is especially important if one chooses to utilize the computational efficiency of SQC methods to study the liquid/vapor interface for instance.\cite{ojha2019time, kaliannan2020impact} The heat of vaporization can be calculated via
\begin{equation*}
  \DHvap{} = \frac{n E_{\ce{H2O}} - \expval{E_{\text{MD}}}}{n} + RT, 
\end{equation*}
where $E_{\ce{H2O}}$ is the energy of an isolated water molecule and $\expval{E_{\text{MD}}}$ is the potential energy of $n$ water molecules averaged during the MD simulation. The \DHvap{} values calculated with various SQC methods and the experimental value are listed in Table~\ref{tab:LiqWatProp}. The original AM1, PM6, and DFTB2 methods throughout predict values for \DHvap{} that are lower than the experimental reference. On the one hand, these results are consistent with their too weakly bound water structures. On the other hand, because liquid water simulated by the \DFTB2iBi{} and \PM6fm{} approaches have a more realistic structure, the values of \DHvap{} are closer to the experimental one. The \AM1W{} model, with its constrained ice-like structure, leads to a highly overestimated value for \DHvap{}. The \GFNxTB{} method, however, is rather a special case. While this method yields a somewhat disordered water structure, it predicts a \DHvap{} that agrees best with the experimental value out of all the the considered SQC methods.  

\subsection{\label{sec:dynamic}Dynamic properties}

\subsubsection{\label{sec:DCoeff}Translational and Rotational Diffusion}

\begin{figure}

\includegraphics[width=0.45\textwidth]{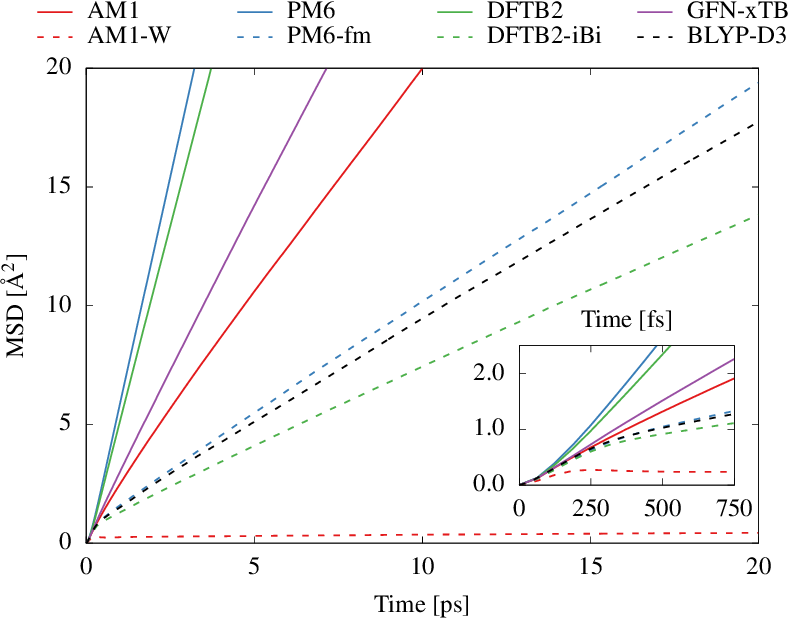}%
\caption{\label{fig:msdW}Mean square displacements of water molecules at ambient conditions as calculated with \BLYPD3{} DFT and different SQC methods. The enclosed inset magnifies the part of the plot near $t=0$.}%
\end{figure}

The translational self-diffusion coefficients $D_\text{PBC}$ listed in Table~\ref{tab:LiqWatProp} were obtained from the slope of the mean square displacement shown in Fig.~\ref{fig:msdW}. It should be noted, however, that due to long-range hydrodynamic effects,\cite{HydrodynPBC, DPBCcorr} the diffusion coefficient is strongly dependent on the size of periodic simulation box causing that $D_{\text{PBC}}$ are a lower bound to the value of an infinite system.\cite{Kuehne2009JCTC} 
With this in mind, the results of Table~\ref{tab:LiqWatProp} immediately show that all original SQC methods yield too high translational diffusion coefficients and hence give a water models that are too fluid.
%\st{We notice also that the \GFNxTB{} water translational self-diffusion is half of that obtained from the DFTB2 method, in agreement with the heat of vaporization being lower for DFTB2} {\color{red} (The problem with the last statement is that it cannot be generalized, I cannot see any general trend in the relation between D and delta $\Delta{H}_{vap}$, presumably because the self-diffusion is rather related to the h-bond free energy, not to any enthalpy or potential energy)}.

For the reparametrized SQC methods \PM6fm{} and \DFTB2iBi{}, respectively, the $D_{\text{PBC}}$ values are in rather good agreement with the experimental value. When the aforementioned finite-size correction is applied,\cite{PM6fm} \PM6fm{} gives a translational self-diffusion value of $2.28\times10^{-5}$~cm$^{2}$/s, which is in outstanding agreement with real water. The diffusion coefficient of \BLYPD3{} DFT is $1.38\times10^{-5}$~cm$^{2}$/s, which is close and below that of \PM6fm{} before finite-size correction. This indicates that \BLYPD3{} DFT can also reproduce the translational dynamics of real water well. The self-diffusion of \DFTB2iBi{} is also expected to be in better agreement with the experimental value when performing a finite-size correction. However, the latter was not included in our simulations using \BLYPD3{} DFT and the \DFTB2iBi{} model, since multiple very long and well-converged runs for various system sizes must be conducted. Finally, \AM1W{} water is in fact not a liquid at all, as already pointed out, and indeed exhibits a mean square displacement that is characteristic of a solid glass.

\begin{figure}
\includegraphics[width=0.45\textwidth]{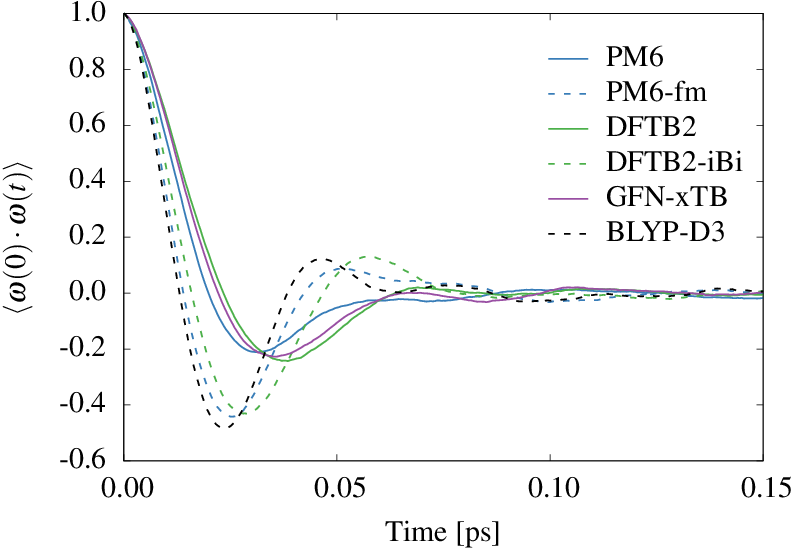}%
\caption{\label{fig:omega_acf}Autocorrelation function of the molecular angular velocity.}%
\end{figure}

In order to characterize the performance of the different SQC methods with respect to the rotational diffusion of water, we have also computed the autocorrelation function (ACF) of the molecular angular velocity, which is shown in Fig.~\ref{fig:omega_acf}. To achieve maximal separation of the rotational and vibrational motions, the angular velocity was calculated in the Eckart frame.\cite{Louck1976} Fig.~\ref{fig:omega_acf} clearly shows that each of the present SQC methods falls into one of two qualitatively distinct classes. On the one hand, we have BLYP-D3 DFT, DFTB2-iBi, and PM6-fm, all of which exhibit a steep decline in their ACF down to $-0.4$ to $-0.5$, and then rebound with positive peak due to the well-known ``caging effect'' of the H-bonds,\cite{Kuehne2009JCTC} before the ACF decays to zero. These are all characteristics of the H-bond network of real water, even though the semi-empirical methods are all redshifted relative to BLYP-D3 DFT (which does provide a good agreement with experiment.\cite{Sharma2005, LowFreq}) On the other hand, we find that PM6, DFTB2, and GFN-xTB severely underestimate the H-bond strength, leading to a considerable redshift in the liberational frequency and, hence, to the observed slow decay of the angular velocity ACF in the time domain. The H-bond network is also too weak to cause a strong rebound peak that we see in the other methods. All these findings can of course be traced back to the potential of mean force depicted in Fig.~\ref{fig:msdW}, with all the methods in the upper row of the figure providing a too low barrier to H-bond bending. %\st{with all the methods in the upper row of Fig.~\ref{fig:msdW} providing a too low barrier --- if at all --- to H-bond bending.}

%{\color{ForestGreen}Andres: As discussed in the email, we proceed without finding the correction for diffussion for DFTB, xTB and BLYP.I computed diffusion for BLYP and found a value of 1.38, close to PM6-fm of 1.54. Text was added.}

\subsubsection{\label{sec:HBDyn}H-bond dynamics}

\begin{figure}
\includegraphics[width=0.45\textwidth]{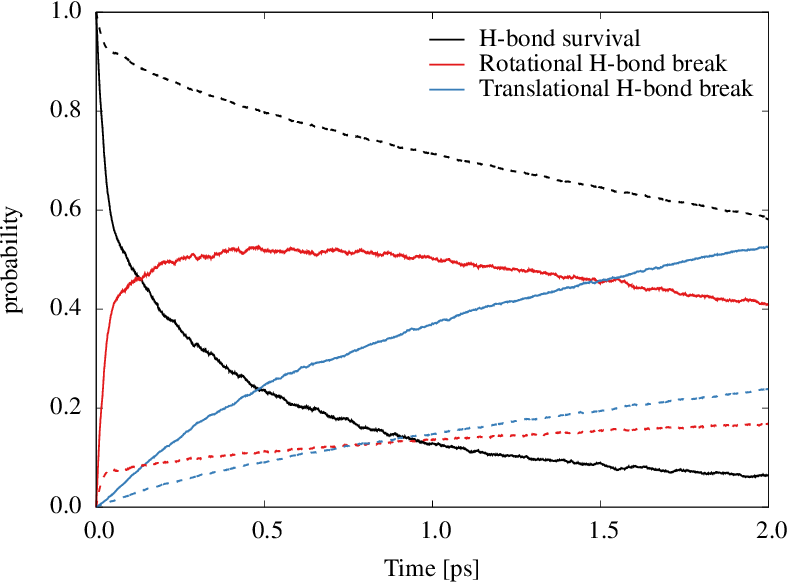}%
\caption{\label{fig:hb_dynamics}The probability that an H-bond that was intact at time $t=0$ is still intact later at time $t$ (black lines). The blue (red) lines show the probability that any H-bond is broken due to relative translational (rotational) motion of the H-bonded pair. Solid lines: GFN-xTB, dashed lines: BLYP-D3 DFT.}%
\end{figure}

The results presented in the preceding section show that, with PM6-fm and AM1-W as exceptions, all the other semi-empirical methods leads to simulated liquid water that is too fluid. Moreover, because each of the methods distorts the two-dimensional potential of mean force in its own way, the impact on rotational and translational diffusion is not the same, leading to an imbalance between both types of diffusion beyond the mere observation that both are enhanced. To clarify this point, we have calculated the survival probability of an H-bond, i.e. the average probability that an H-bond still exists at some time $t$, given that it was intact at $t=0$. For the fraction of H-bonds that are broken at any time $t$, we further calculate the probability that it was broken due to translational (H-bond stretching), or rotational diffusion (H-bond bending).\cite{luzar1996hydrogen, luzar1996effect} Fig.~\ref{fig:hb_dynamics} shows the three probabilities (H-bond survival, translational H-bond breaking, and rotational H-bond breaking) for the GFN-xTB and BLYP-D3 DFT methods, respectively. Our choice of GFN-xTB here is based on the observation that it outperforms all the other methods in predicting the heat of vaporization. Although GFN-xTB overestimates the translational diffusion coefficient almost twofold, the distortion of the H-bond dynamics due to the enhanced rotational dynamics is even worse, with 40\% of the H-bonds being broken within 50~fs just by the initial liberational oscillation alone. The contribution of translational diffusion to H-bond breaking only catches up with that of the rotational diffusion after 1.5~ps (in contrast to 0.75~ps for BLYP-D3 DFT). It would have actually taken a longer time had it not been for the reformation of some of the rotationally broken H-bonds, which consequently leads to a decrease in the rotational contribution to H-bond breaking at longer times.

%{\color{red}Q: should we move the figure from SI to article?}
%{\color{blue} XinWu: I've moved it from SI to here.}

\section{\label{sec:fazit}CONCLUSIONS}

Historically, the emergence of SQC methods was driven by a very pragmatic philosophy: the need to
compute molecular properties in an era when \textit{ab initio} quantum mechanical approaches were
too complicated to be applied, even for small molecules.\cite{pople1965xndo} % Let's cite an old paper.
The promise is that through approximations combined with careful parametrization, one can gain substantial savings in computational effort and possibly even some improvement in the accuracy. Nowadays, with the feasibility of very accurate correlated electronic structure calculations on small-sized systems,\cite{harding2008high} one of the possible niches of SQC methods are high-throughput calculations on large system sizes and, in case of condensed phase dynamics, applications involving length and time scales that are beyond the reach of DFT-based AIMD.\cite{schade2022towards, schade2023breaking}
In this work, we have critically analyzed the performance of some of the most popular NDDO-, as well as
DFTB-type SQC methods regarding their accuracy in predicting static and dynamical properties of bulk liquid water at ambient conditions through MD simulations. With the exception of the PM6-fm model, which performs rather well, all of the investigated SQC methods are rather limited in their performance to predict the most fundamental static and dynamic properties and are not suitable for simulations of liquid water. Remarkably, with the exception of the PM6-fm and AM1-W approaches (the latter gives a solid glass), all of the SQC methods commonly predict hydrogen bonds that are too weak, leading with variable extents to a less structured liquid that is extremely fluid, thus grossly overestimating the translational and/or rotational diffusion coefficients, thereby leading to a highly distorted H-bond dynamics. An intriguing question that needs to be addressed in future works is whether this common misbehavior of different SQC methods is because of some deficiencies in the underlying theory, or due to the parametrization strategy that is focused on various properties of small water clusters of varying sizes. Regardless of the answer to this question, the substantially improved performance of the PM6-fm model in comparison to the original PM6 shows that these shortcomings can be remedied through specific parametrization. In light of these findings, we conclude that the PM6-fm model is an efficient, low-cost alternative to DFT for simulating liquid water.

%XinWu: I'm not sure if this is also a conclusion: The specific parametrization can be used as a powerful tool to improve the accuracy of an SQC method for a system of interest. 

%{\color{red}(Is the last statement even correct? I am not familiar with the parametrization strategy of everyone of this array of methods. Xin?)}. 

%{\color{blue}XinWu: Just comment (not conclusion): the NDDO-type as well as the DFTB-type SQC methods have theoretical flaws, e.g. a huge amount of significant electronic integrals are neglected. To compromise these deficiencies, all SQC methods are \emph{trained} by parametrization, e.g. different sets of molecules and/or different optimization algorithms. It's very difficult to quantify the impact of the underlying theory and parametrization. Concerning liquid water as we can see, although the PM6 method performs bad, the reparametrized PM6-fm method is really good.}

\begin{acknowledgments}
The authors gratefully acknowledge the Gauss Centre for Supercomputing e.V. (www.gauss-centre.eu) for funding this project by providing computing time through the John von Neumann Institute for Computing (NIC) on the GCS Supercomputer JUWELS at Jülich Supercomputing Centre (JSC). The generous allocation of computing time on the FPGA-based supercomputer ``Noctua'' by the Paderborn Center for Parallel Computing (PC$^2$) is kindly acknowledged. This project has received funding from the European Research Council (ERC) under the European Union's Horizon 2020 research and innovation programme (Grant Agreement No. 716142). T.D.K. and C.P. kindly acknowledges funding from Paderborn University's research award for ``GreenIT'' A.~H.\ thanks the Alexander von Humboldt Foundation for his postdoctoral research fellowship, whereas H.~E. is grateful for funding (grant EL 815/2) from the German Research Foundation (DFG). 
\end{acknowledgments}

\bibliography{water}

\end{document}